\documentclass[cpp,11pt]{wmthesis}
\usepackage{graphicx}
\usepackage{color}
\usepackage{url}
\usepackage{pstricks}
\usepackage{epsfig}
\usepackage{verbatim}
\usepackage[subfigure]{tocloft}
\usepackage{latexsym}
\usepackage{rotating}
\usepackage{multirow}
\usepackage[numbers]{natbib}
\usepackage{mdwlist}
\usepackage[algochapter]{algorithm2e}
\usepackage{subfigure} 
\usepackage{tikz}
\newcommand*\circled[1]{\tikz[baseline=(char.base)]{
            \node[shape=circle,draw,inner sep=0.5pt] (char) {#1};}}

\def\BEGINITEMIZE{\begin{itemize}}
\def\ENDITEMIZE{\end{itemize}}

\def\defaultpenalty{1000} \clubpenalty=\defaultpenalty
\widowpenalty=\defaultpenalty

\newcommand{\inputfig}[2][\empty]{ 
   \begin{center} %
   	\ifx\empty#1 \includegraphics{../figs/#2}
	\else\scalebox{#1}{\includegraphics{../figs/#2}}\fi
   \end{center}}

\renewcommand\listfigurename{\begin{center}\Large\normalfont LIST OF FIGURES\vspace{-.35in}\end{center}}
\renewcommand\listtablename{\begin{center}\Large\normalfont LIST OF TABLES\vspace{-.35in}\end{center}}

\begin{document}
\doublespacing

\thesisTitle{Enhancing Bug Reports for Mobile Apps}
\thesisAuthor[Kevin Patrick Moran]{Kevin Patrick Moran}
\thesisMonth{August}
\thesisYear{2015}
\thesisAdvisor{Associate Professor Denys Poshyvanyk}
\thesisLocation{Maitland, FL}
\thesisDegreeOne{Bachelor of Science, College of the Holy Cross, 2013}
\thesisCommittee[Computer Science]{\ThesisAdvisor}
\thesisCommittee[Computer Science]{Associate Professor Gang Zhou}
\thesisCommittee[Computer Science]{Professor Xu Liu}

\thesisDedication{In loving memory of Thomas P. Moran, Mary E. Lavin, and Paul B. Lavin; your lives inspired many people -- myself included -- to accomplish great things.}

\thesisAcknowledge{acknowledge}
\thesisAbstract{abstract}

\makeThesisProlog

\chapter{Introduction}

\section{Motivation} 
\label{Intro:motivation}

	 Smartphones and mobile computing have skyrocketed in popularity in recent years, and adoption has reached near-ubiquitous levels with over 2.7 billion active smartphone users in 2014 \cite{24MobilityReport}.  An increased demand for high-quality, robust mobile applications is being driven by a growing user base that performs an increasing number of computing tasks on ``smart'' devices. Due to this demand, the complexity of mobile applications has been increasing, making development and maintenance challenging. The intense competition present in mobile application marketplaces like Google Play and the Apple App Store, means that if an app is not performing as expected,  due to bugs or lack of desired features, 48\% of users are less likely to use the app again and will abandon it for another one with similar functionality \cite{app-abandonment}. 
	
	Software maintenance activities are known to be generally expensive and challenging \cite{25Tassey:NIST}.  One of the most important maintenance tasks is bug report resolution.  However, current bug tracking systems such as Bugzilla \cite{bugzilla}, Mantis \cite{mantis}, the Google Code Issue Tracker \cite{google-code}, the GitHub Issue Tracker \cite{github-it}, and commercial solutions such as JIRA \cite{jira} rely mostly on unstructured natural language bug descriptions.  These descriptions can be augmented with files uploaded by the reporters (e.g., screenshots). As an important component of bug reports, reproduction steps are expected to be reported in a structured and descriptive way, but the quality of description mostly depends on the reporter's experience and attitude towards providing enough information. Therefore, the reporting process can be cumbersome, and the additional effort means that many users are unlikely to enhance their reports with extra information \cite{11Bettenburg:MSR08,31Davies:ESEM2014,32Bettenburg:ICSM08, 34Aranda:ICSE09}. 
	
	 A past survey of open source developers conducted by Koru et al. has shown that only  $\approx$ 50\% of developers believe bug reports are always complete \cite{33Koru:IEEE2004}.  Previous studies have also shown that the information most useful to developers is often the most difficult for reporters to provide and that the lack of this information is a major reason behind non-reproducible bug reports \cite{4Joorabchi:MSR14, 3Bettenburg:FSE08}.   Difficulty providing such information, especially reproduction steps, is compounded in the context of mobile applications due to their complex event-driven and GUI-based nature. Furthermore, many bug reports are created from textual descriptions of problems in user reviews.  According to a recent study by Chen et al. \cite{Chen:icse2014}, only a reduced set of user reviews can be considered useful and/or informative. Also, unlike issue reports and development emails, reviews do not refer to details of the app implementation.
	
	The above issues point to a more prominent problem for bug tracking systems in general: the \textit{lexical gap} that normally exists between bug reporters (e.g., testers, beta users) and developers.  Reporters typically only have functional knowledge of an app, even if they have development experience themselves, whereas the developers working on an app tend to have intimate code level knowledge.  In fact, a recent study conducted by Huo et al. corroborates the existence of this knowledge gap as they found there is a difference between the way experts and non-experts write bug reports as measured by textual similarity metrics \cite{35Huo:ICSME14}.  When a developer reads and attempts to comprehend (or reproduce) a bug report, she has to bridge this gap, reasoning about the code level problems from the high-level functional description in the bug report.  If the lexical gap is too wide the developer may not be able to reproduce and/or subsequently resolve the bug report.

\section{Contributions}
\label{Contributions}

	  To address this fundamental problem of making bug reports more useful (and reproducible) for developers, we introduce a novel approach, which we call FUSION, that relies on a novel \textit{Analyze $\rightarrow$ Generate} paradigm to enable the auto-completion of Android bug reports in order to provide more actionable information to developers.  In the context of this work, we define auto-completion as suggesting relevant actions, screen-shots, and images of specific GUI-components to the user in order to facilitate reporting the steps for reproducing a bug.  FUSION first uses fully automated static and dynamic analysis techniques to gather screen-shots and other relevant information about an app before it is released for testing. Reporters then interact with the web-based report generator using the auto-completion features in order to provide the bug reproduction steps.  By linking the information provided by the user with features extracted through static and dynamic analyses, FUSION presents an augmented bug report to the developer that contains immediately actionable information with well-defined steps to reproduce a bug. The work presented in this thesis represents an extension of a published conference paper at FSE'15 \cite{Moran:FSE15}.
	
	We evaluate FUSION in a study comparing bug reports submitted using our system to the bug reports produced using Google Code Issue Tracker, involving 28 participants, reporting bugs for 15 real-world failures stemming from 14 open source Android apps. 
	
	Our paper makes the following noteworthy contributions:
We evaluate FUSION in a study comparing bug reports submitted using our system to bug reports produced using Google Code Issue Tracker, involving 28 participants reporting bugs for 15 real-world failures stemming from 14 open source Android apps. 
	
Our paper makes the following noteworthy contributions:
\begin{enumerate}
\item We design and implement a novel approach for auto-completing and augmenting Android bug reports, called FUSION, which leverages static and dynamic analyses, and provides actionable information to developers. The tool facilitates the reporting, reproduction and subsequent resolution of Android bugs. The program analysis techniques of the apps can be run on \textit{both} physical devices and emulators;
\item We design and carry out a comprehensive user study to evaluate the \textit{user experience} of our approach and the \textit{quality} of bug reports generated using FUSION compared to the Google Code Issue Tracker. The results of this study  demonstrate that FUSION enables developers to submit bug reports that are more likely to be reproducible compared to reports written entirely in natural language;  
\item We make FUSION and all the data from the experiments available for researchers \cite{appendix} in hope that this work spurs new research related to improving the quality of bug reports and bug reporting systems.  
\end{enumerate}			
\chapter{Related Works}
\label{related_works}

 Bug and error reporting has been an active area of research in the software engineering community.  However, little work has been conducted to improve the lack of structure in the reporting mechanism for entering reproduction steps, and adding corresponding support in bug tracking systems. Therefore, in this section, we briefly survey the features of current bug reporting systems and the studies that motivated this work.  We outline the current types of information that bug reporting systems attempt to elicit from users and explain how FUSION improves upon this method for collecting information.  Then we differentiate our work from approaches for reproducing in-field failures and explain how our work compliments existing research on bug reporting. 

\section{Existing Bug Reporting Systems}
\label{bug_systems}

\begin{figure*}[tb]
\centering
\includegraphics[width=\linewidth]{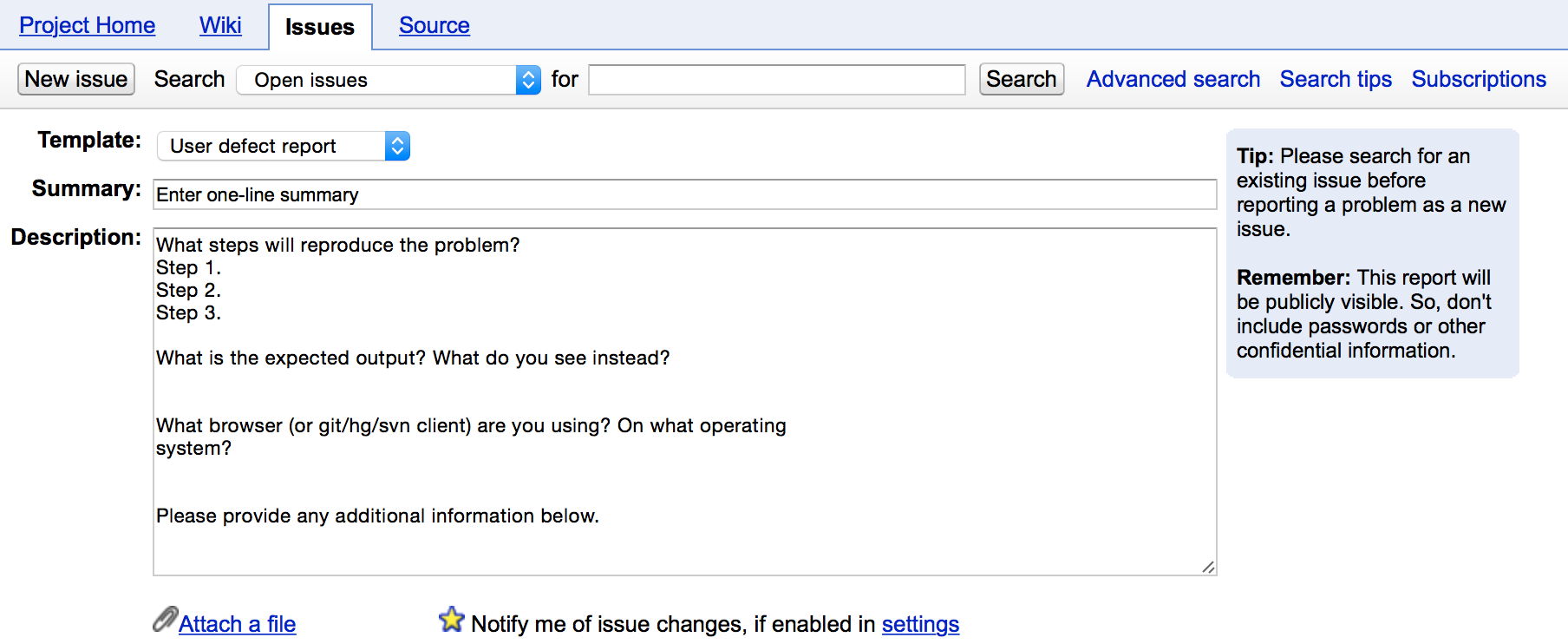}
\vspace{-0.5cm}
\caption{\textbf{The Google Code Issue Tracker:} An example of a popular Issue Tracking System with semi-structured fields for reporters to construct issue tickets. Image taken from \cite{google-code}}
\label{gcit}
\vspace{-0.3cm}
\end{figure*}

\begin{figure*}[b]
\centering
\includegraphics[width=0.7\linewidth]{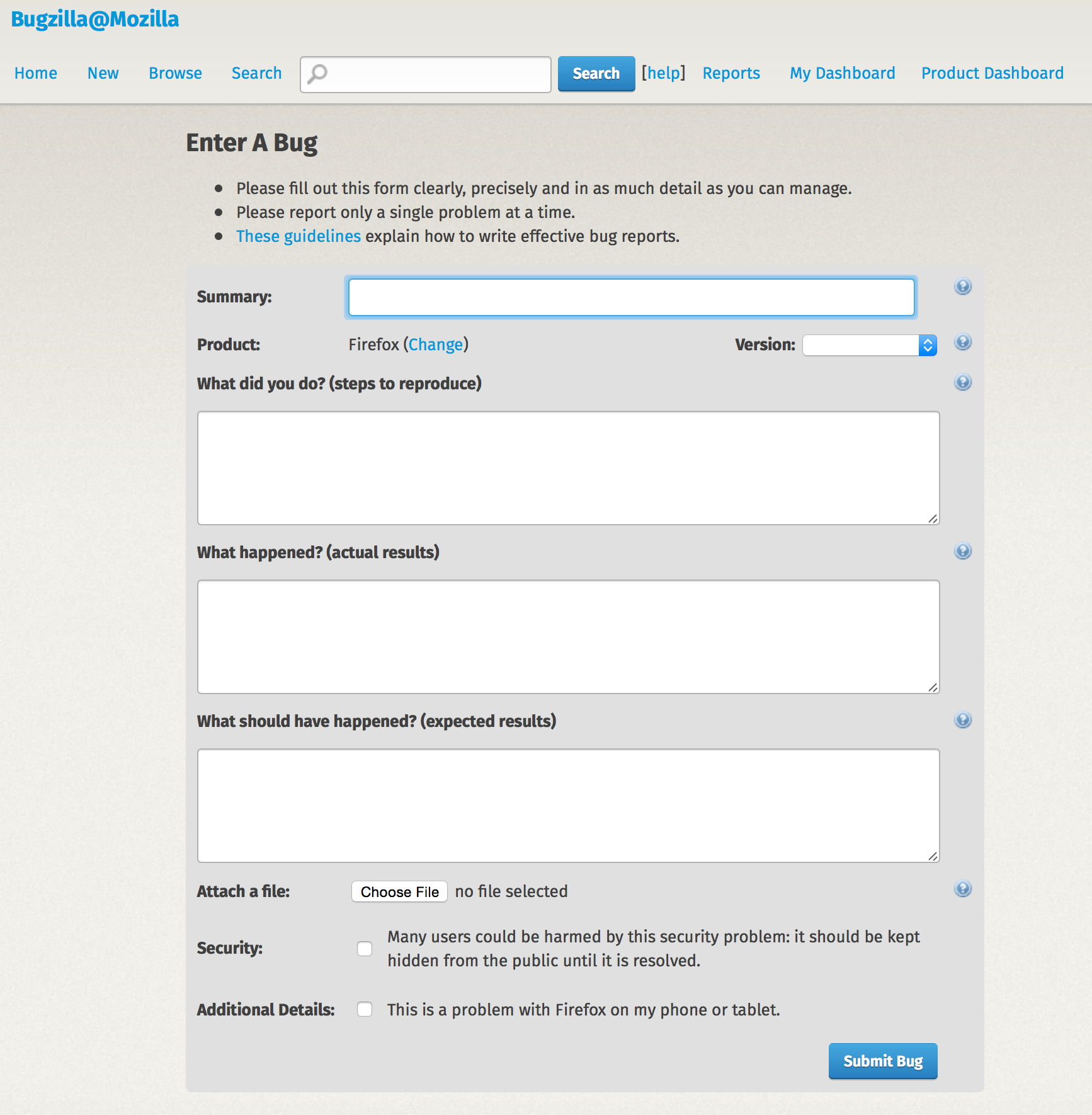}
\vspace{-0.3cm}
\caption{\textbf{Bugzilla Issue Tracker:} An example of a popular Issue Tracking System with separate fields that prompts users to enter reproduction steps and expected/actual results of the steps. Image taken from \cite{bugzilla}}
\label{bugzilla}
\vspace{-0.3cm}
\end{figure*}

\begin{figure*}[tb]
\centering
\includegraphics[width=\linewidth]{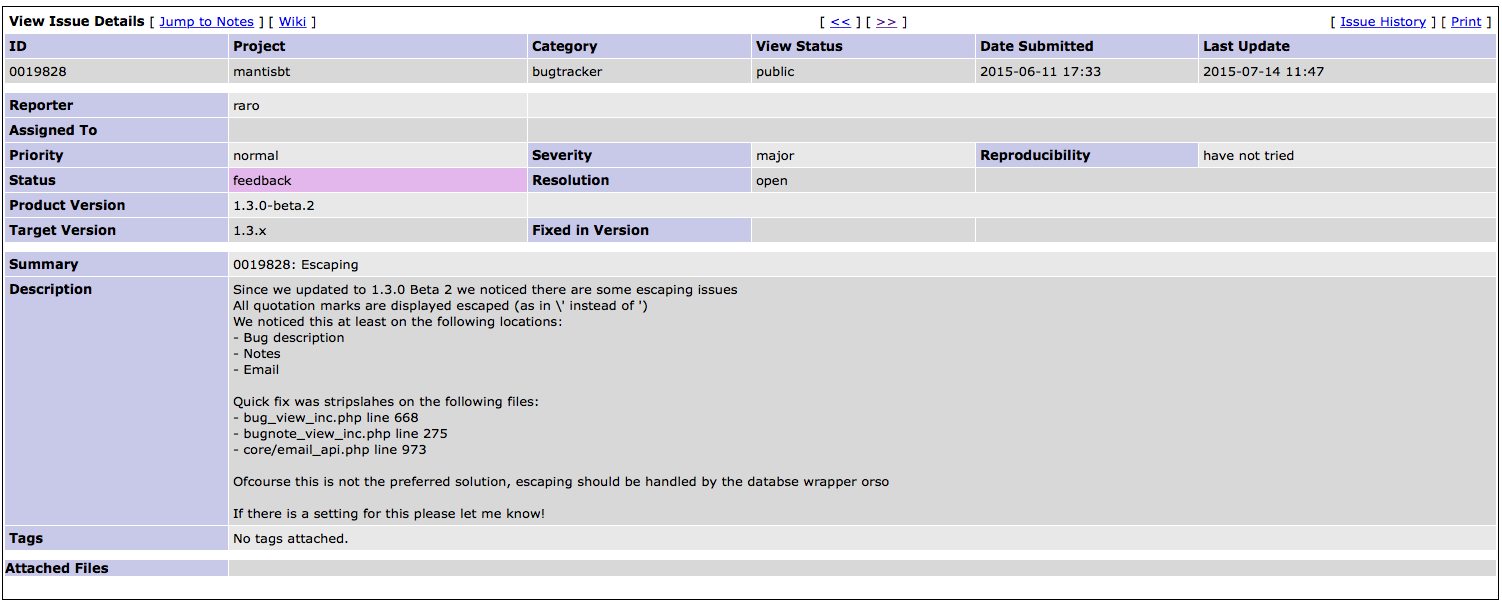}
\vspace{-0.3cm}
\caption{\textbf{Mantis Issue Tracker:} An example of a popular Issue Tracking System containing a large amount of contextual information. Image taken from \cite{mantis}}
\label{mantis}
\vspace{-0.3cm}
\end{figure*}

The purpose of a bug reporting system, sometimes referred to as an issue tracker, is twofold: First, such systems must provide a coherent and easy to use mechanism for reporters to accurately and completely describe a software defect or feature addition/enhancement.  Second, they must organize this information and present it to the developer in a meaningful fashion.  Most current bug reporting systems rely upon unstructured natural language descriptions in their reports.  However, some systems do offer more functionality.  For instance, the Google Code Issue Tracker (GCIT) \cite{google-code} (See Figure \ref{gcit}) offers a semi-structured area where reporters can enter reproduction steps and expected input/output in natural language form (i.e., the online form asks: "What steps will reproduce the problem?").  Nearly all current issue trackers offer structured fields to enter information such as tags, severity level, assignee, fix time, and product/program specifications.  Some web-based bug reporting systems (e.g. Bugzilla \cite{bugzilla} (See Figure \ref{bugzilla}), Jira \cite{jira}, Mantis \cite{mantis} (See Figure \ref{mantis}), UserSnap \cite{usersnap}, BugDigger  \cite{22BugDigger}) facilitate reporters including screenshots.  One commonality that most of these reporting systems share is that reporters and developers share the same view of the bug report report.  That is, there is no differentiation between the information and view of the issue report that reporter sees and that which the developer sees. This is yet another example of how current issue tracking systems do not effectively handle the lexical gap that exists between developers and testers/reporters.  Ideally, the user-facing reporting mechanism should be familiar and easy to use, and the developer report should be detailed and suitable for the maintenance task at hand (e.g. feature addition/enhancement, bug fixing). In other words, the issue tracker system, \textbf{not the developer} should bridge the \textit{lexical knowledge gap} that typically exists between reporters and developers.  This is precisely what FUSION accomplishes.  By leveraging information gleaned from program analysis, FUSION is able to present to the reporter the familiar interface of a mobile application GUI, and suggest the steps to reproduce a bug.

\section{Bug Reporting Studies}
\label{bug_studies}

The problem facing many current bug reporting systems is that typical natural language reports capture a coarse grained level of detail that makes developer reasoning about defects difficult. This highlights the underlying \textit{task} that bug reporting system must accomplish: \textit{bridging the lexical knowledge gap between typical reporters of a bug and the developers that must resolve the bugs.}  In order for an issue tracking system to effectively accomplish this task, it must facilitate the entry of certain types of crucial information that developers find useful.  
	Previous studies on bug report quality and developer information needs highlight several factors that can impact the quality of bug reports \cite{15Breu:CSCW10, 4Joorabchi:MSR14, 3Bettenburg:FSE08}:

\begin{itemize}
	\item Other than ``Interbug dependencies'' (i.e., a situation where a bug was fixed in a previous patch), \textit{insufficient information} in bug reports is one of the leading causes of non-reproducible bug reports \cite{4Joorabchi:MSR14};
	\item Developers consider (i)\textit{steps to reproduce}, (ii)\textit{stack traces}, and (iii)\textit{test cases/scenarios} as the most helpful sources of information in a bug report \cite{3Bettenburg:FSE08};
	\item Information needs are greatest early in a bug's life cycle, therefore, a way to easily add the above features is important during bug report creation \cite{15Breu:CSCW10}.
\end{itemize}

Using these issues as motivation, we developed FUSION with two major goals in mind: (i) \textit{provide bug reports to developers with immediately actionable knowledge (reliable reproduction steps)} and (ii) \textit{facilitate reporting by providing this information through an auto-completion mechanism.}

It is worth noting that one previous study conducted by Bhattacharya et$.$ al$.$ \cite{5Bhattacharya:CSMR13} concluded that most Android bug reports for open source apps are of high-quality, however in their study only $\approx$ 46\% of bug report contained steps to reproduce, and and even lesser amount ($\approx$ 20\%) contained additional information (e.g. bug-triggering input or even an app version).  Therefore, there is clearly room for improvement in terms of the type of information that is contained within open source Android bug reports. By helping auto-complete the reproduction steps using guided suggestions for reporter GUI actions and corresponding components, we facilitate the reporter providing this information in the bug report which is both useful from a developer's  perspective, and typically difficult to provide from a reporter's perspective.

\section{In-Field Failure Reproduction}
\label{field_failures}

A body of work known as in-field failure reproduction \cite{27Bell:ICSE13, 26Jin:ISSTA13, 8Zhou:ICSE12, 29Clause:ICSE07,18Jin:ICSE12,41Artzi:ECOOP2008,50Kifetew:ICST2014,43Cao:ASE14} shares similar goals with our approach. These techniques collect run-time information (e.g., execution traces) from instrumented programs that provide developers with a better understanding of the causes of an in-field failure, which will subsequently help expedite the fixing of those failures.  However, there are several key differences that set our work apart and illustrate how FUSION improves upon the state of research.   

	\textit{First}, techniques regarding in-field failure reproduction rely on potentially expensive program instrumentation, which requires developers to modify code, and introduce overhead.  FUSION is completely automatic, our static and dynamic analysis techniques only need to be applied once for the version of the program that is released for testing.  Furthermore, the analysis process can be done without the need for instrumentation of programs in the field.  \textit{Second}, current in-field failure reproduction techniques require an oracle to signify when a failure has occurred (e.g., a crash).  FUSION is not an approach for crash or failure detection, it is designed to support testers during the bug reporting process. \textit{Third}, these techniques have not been applied to mobile apps and would most likely need to be optimized further to be applicable for the corresponding resource-constrained enviornment.  
	
\section{Bug and Error Reporting Research}
\label{bug_research}

A subset of prior work has been focused on bug and crash triage \cite{7Shokripour:MSR13, 10Naguib:MSR13,40Jeong:FSE2009,44Kim:TOSE2013,45Park:AAAI2011,46Kim:DSN2011,51Haihao:ICST2011,34Aranda:ICSE09,Linares-Vasquez:ICSM2012,Menzies:ICSM2008,Gethers:ASE11,Hossen:ICPC2014,Kagdi:ICPC09,Huzefa:JSEP12}.  The techniques associated with this topic typically employ different program analysis and machine learning or natural language processing techniques to match bug reports with appropriate developers.  Our proposed research compliments developer recommendation frameworks, as FUSION can provide these frameworks with more detailed ``knowledge'' than current state of practice bug reporting systems.  

	A significant amount of research has been conducted concerning the summarization \cite{1Mani:FSE12,11Bettenburg:MSR08,20Rastkar:ICSE10,33Koru:IEEE2004,53Weiss:MSR2007,Czarnecki:ICSM2012}, fault localization \cite{8Zhou:ICSE12,13Wang:ICPC14,37Rahman:FSE2011,38Baudry:ICSE2006,39Vidacs:CSMR14,42Wu:ISSTA2014,47Marsi:STVR2010,49Ayewah:IS2008,52Cleve:ICSE2005,54Dallmeier:LNCS2005}, classification and detection of duplicate bug reports \cite{4Joorabchi:MSR14, 14Nguyen:ASE12, 17Wang:ICSE08, 19Guo:ICSE10,  21Zhou:CIKM12, 36Gu:ICSE10,48Podgurski:ICSE2003}. Research on these topics is primarily concerned with duplicate bug report detection, localizing bug reports to specific areas of source code, and effectively summarizing reports for developers with the most pertinent information.  
	
	Again, the work presented in this paper compliments these categories of research as bug reports created with FUSION can provide more detailed information, easliy linking the bug back to source code, allowing for better localization, summarization and, potentially, duplicate detection.  It is worth noting that work by Bettenburg et$.$ al$.$ on extracting structural information from bug reports is also related, however, we aim at helping auto-complete the structured reproduction steps at the time of report creation, rather than extracting it after the fact \cite{11Bettenburg:MSR08}. 

\chapter{The FUSION Approach}
\label{approach}

\begin{figure}[p]
\centering
\includegraphics[width=0.74\textwidth]{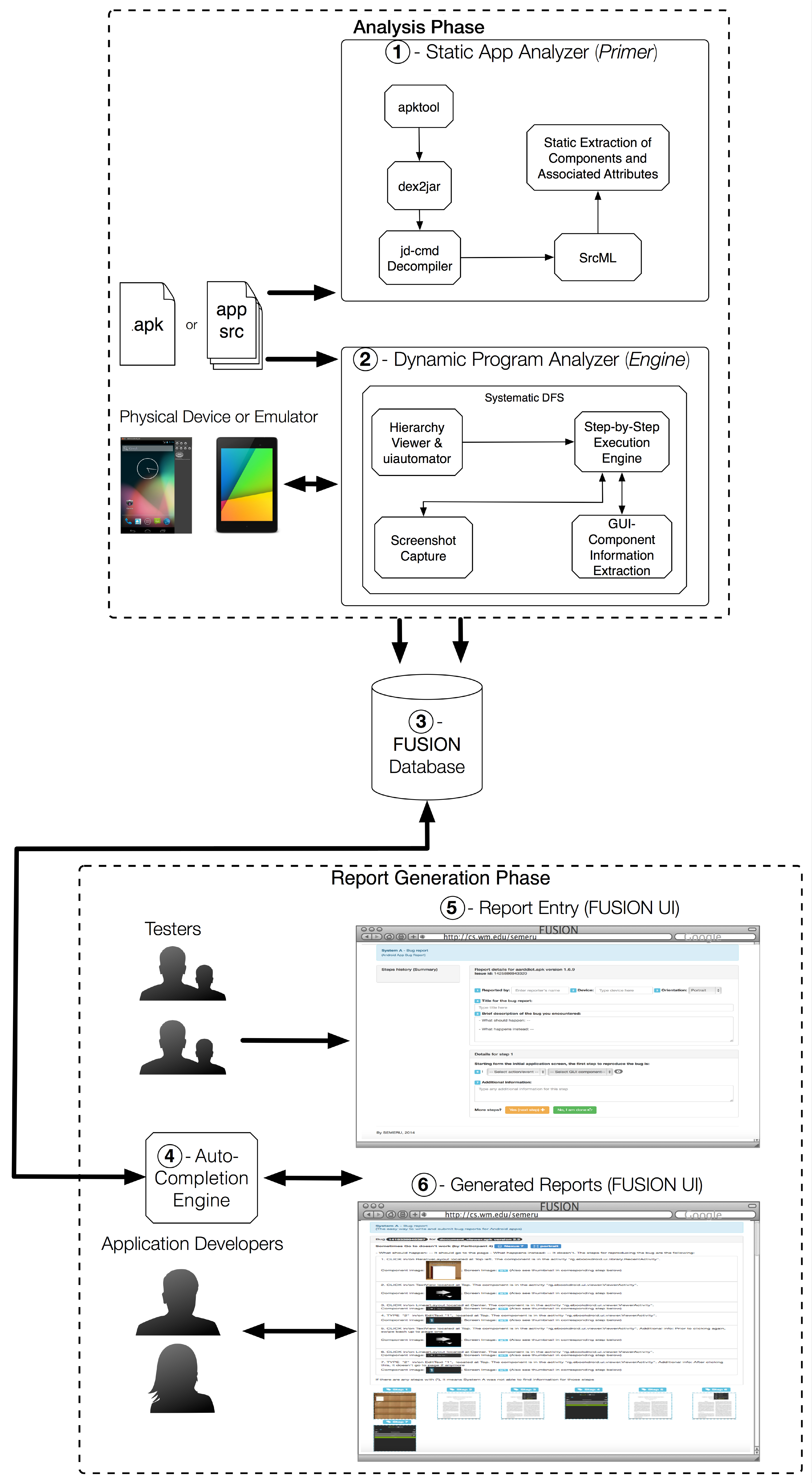}
\vspace{-0.5cm}
\caption{\textbf{Overview of FUSION Workflow:} First static and dynamic app analysis is performed on the target app, then the Auto-Completion Engine uses the information gleaned by the analyses in order to help the user auto-complete reproduction steps of a bug for the target app.}
\label{Design}
\vspace{-0.3cm}
\end{figure}

FUSION's \textit{Analyze} $\rightarrow$ \textit{Generate} workflow corresponds to two major phases. In the \textit{Analysis Phase} FUSION collects information related to the GUI components and event flow of an app through a combination of static and dynamic analysis.  Then in the \textit{Report Generation Phase} FUSION takes advantage of the GUI centric nature of mobile apps to both auto-complete the steps to reproduce the bug and augment each step with contextual application information.  The overall design of FUSION can be seen in Figure \ref{Design}.  We encourage readers to view videos of our tool in use, complete with commentary that are available at our online appendix \cite{appendix} The key idea behind the FUSION workflow is: \textit{ program analysis, performed preemptively before an app is released for testing, can be used as a means to aid reporters to easily provide information that developers need during the bug reporting process to reproduce and fix app issues}.

\section{Analysis Phase}
\label{analyze_phase}

 The \textit{Analysis Phase} collects all the information required for the \textit{Report Generation Phase} operation. The first phase has two major components: 1) static analysis \emph{(Primer)}, and 2) dynamic program analysis \emph{(Engine)} of a target app. The information generated by \emph{(Primer)} and \emph{(Engine)} is required by the Report Generation Phase. The \textit{Analysis phase} must be performed before each version of an app is released for testing or before it is published to end users.  Both components of the \emph{Analysis Phase} store their extracted data in the FUSION database (Fig. \ref{Design} - \circled{3}).

\subsection{Static Analysis (Primer)}
\label{primer}

The goal of the \emph{Primer} (Fig. \ref{Design} - \circled{1})
   is to extract all of the GUI components and associated information from the app source code.   For each GUI component, the \emph{Primer} extracts: (i) possible actions on that component, (ii) type of the component (e.g., Button, Spinner), (iii) activities the component is contained within, and (iv) class files where the component is instantiated.  
    Thus, this phase gives us a universe of possible components within the domain of the application, and establishes traceability links connecting GUI components that reporters operate upon to code specific information such as the class or activity they are located within.  
    
    The \emph{Primer} is comprised of several steps to extract the information outlined above. First it uses the  {\tt dex2jar}\cite{dex2jar} and  {\tt jd-cmd} \cite{jd-cmd}  tools for decompilation, then, it converts the source files to an XML-based representation using {\tt srcML} \cite{srcml}. We also use {\tt apktool} \cite{apktool} to extract the resource files from the app's APK.  The {\tt id}s, and types of GUI components were extracted from the xml files located in the app's resource folders (\textit{i.e.},  {\tt /res/layout} and  {\tt /res/menu} of the decompiled application or src).  Using the {\tt srcML} representation of the source code we are able to parse and link the GUI-component information to extracted app source files.  

\subsection{Dynamic Analysis (Engine)}
\label{engine}

The \emph{Engine} (Fig. \ref{Design} - \circled{2}) is used to glean dynamic contextual information, such as the location of the GUI component on the screen, and enhance the database with both run-time GUI and application event flow information. The goal of the \emph{Engine} is to explore an app in a systematic manner ripping and extracting run-time information related to the GUI components during execution including: (i) the text associated with different GUI components (e.g., the ``Send'' text on a button to send an email message), (ii) whether the GUI component triggers a transition to a different activity, (iii) the action performed on the GUI component during systematic execution, (iv) full screen-shots before and after each action is performed, (v) the location of the GUI component object on the test device's screen, (vi) the current Activity and window of each step, (vii) screen-shots of the specific GUI component, and (viii) the object index of the GUI component (to allow for differentiation between different instantiations of the same GUI component on one screen).

    The \emph{Engine} performs this systematic exploration of the app using the {\tt UIAutomator} \cite{uiautomator} framework included in the Android SDK.  This systematic execution of the app is similar to existing approaches in GUI ripping \cite{55Amalfitano:ASE2012,Takala:ICST2011,Ravindranath:Mobisys2014,Amalfitano:ASE2012,Azim:OOPSLA2013,Machiry:FSE2013,Choi:OOPSLA2013,Nguyen:TOSE2014}. Using the {\tt UIAutomator} framework allows us to capture cases that are not captured in previous tools such as pop-up menus that exist within menus, internal windows, and the onscreen keyboard.  To effectively explore the application we implemented our own version of a systematic depth-first search (DFS) algorithm for application traversal that performs click events on all the clickable components in the GUI that can be reached using the DFS-based traversal.
      
    During the ripping, before each step is executed on the GUI, the \emph{Engine} makes a call to {\tt UIAutomator} subroutines to extract the contextual information outlined above regarding each GUI component displayed on the device screen.  We then execute the action associated with each GUI component in a depth-first manner on the current screen. Our current implementation of DFS only handles the click/tap action, however, as this is the most common action, we are still able to explore a significant amount of an application's functionality.  
     
     In the DFS algorithm, if a link is clicked that would normally transition to a screen in an external activity (e.g., clicking a web link that would launch the Chrome web browser app) we execute a \textit{back} command in order to stay within the current app.  If the DFS exploration exits the app to the home screen of the device/emulator for any reason, we simply re-launch the app and  continue the GUI traversal.  During the DFS exploration, the \emph{Engine} captures each activity transition that occurs after each action is performed (e.g., whether or not a new activity is started/resumed after an action to launch a menu).  This allows FUSION to build a model of the app execution that we will later use to help track a reporter's relative position in the app when they are using the system to record the steps to reproduce the bugs.

\section{Report Generation Phase}
\label{generation_phase}

We had two major goals when designing the \textit{Report Generation Phase} component of FUSION:
    
\begin{enumerate}
  \item Allow for traditional natural language input in order to give a high-level overview of a bug.
  \item Auto-complete the reproduction steps of a bug through suggestions derived by tracking the position of the reporter's step entry in the app event flow.
\end{enumerate}
       
                During the \emph{Report Generation Phase} FUSION aids the reporter in constructing the steps needed to recreate a bug by making suggestions based upon the ``potential" GUI state reached by the declared steps.  This means for each step $s$, FUSION infers --- online --- the GUI state $GUI_s$ in which the target app should be, by taking into account the history of steps.   For each step, FUSION verifies that the suggestion made to the reporter is correct by presenting the reporter with contextually relevant screen-shots, where the reporter selects the screen-shot corresponding to the current action the reporter wants to describe. 
                        
\subsection{Report Generator User Interface}
\label{user_interface}

 \begin{figure}[p]
\centering
\includegraphics[width=\linewidth]{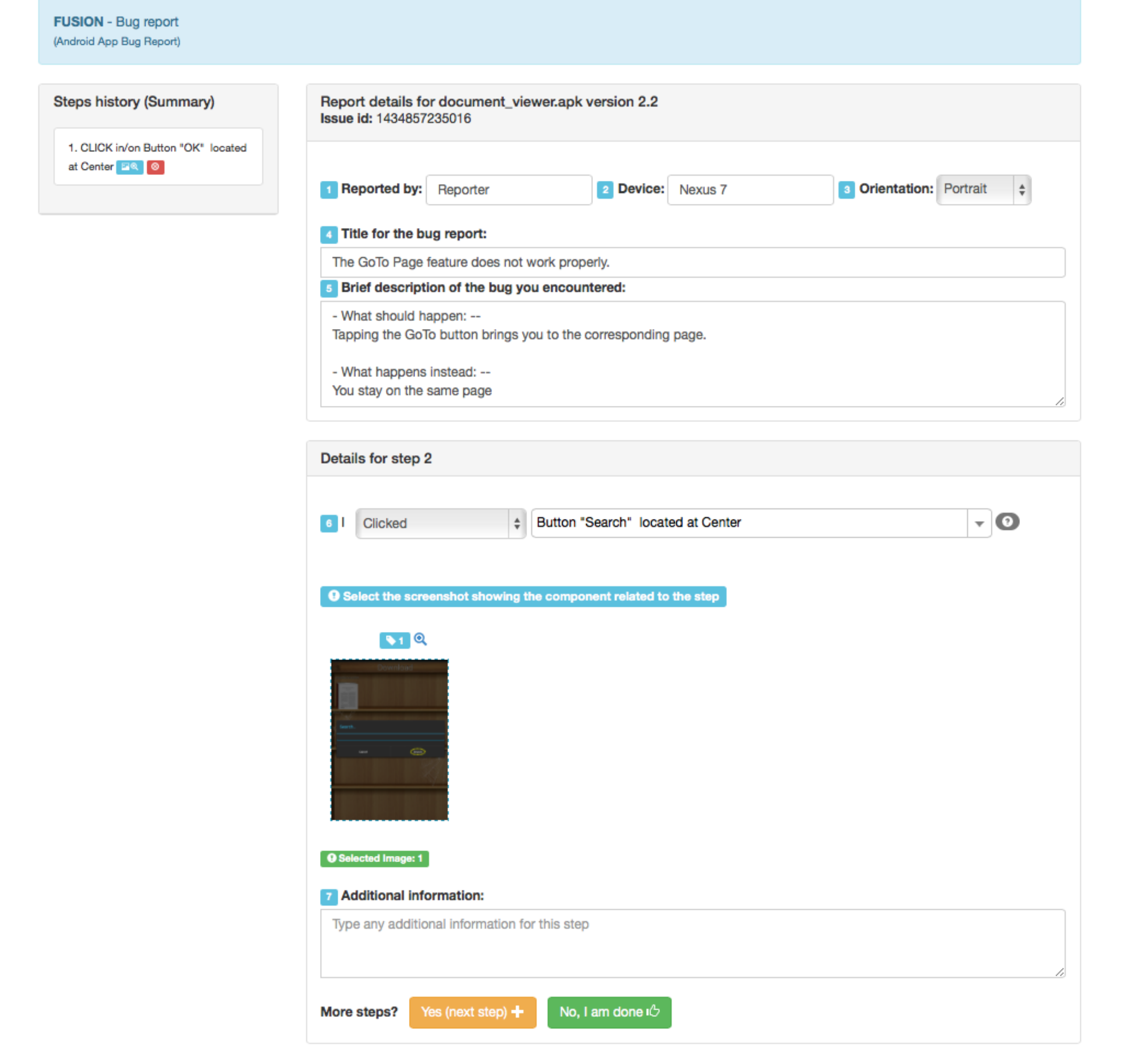}
\vspace{-0.8cm}
\caption{\textbf{FUSION Reporter Interface:} This figure shows the FUSION reporter web interface, with an area for contextual information and a natural language description of the bug at the top of the page, an area for reporting reproduction steps, and an area showing the history of the steps entered with an option to view or delete past steps.}
\label{front_end}

\end{figure}

    After first selecting the app to report an issue for, a reporter interacts with FUSION by filling in some brief contextual information (i.e., name, device, title) and a brief textual description of the bug in question in the top half of the UI.  Next, the reporter inputs the steps to reproduce the bug using the auto-completion boxes in a step-wise manner, starting from the initial screen of a cold app launch\footnote{Cold-start means the first step is executed on the first window and screen displayed directly after the app is launched}, and proceeds until the list of steps to reproduce the bug is exhausted.   Let us consider a running example where the user is filling out a report for the Document Viewer bug in Table \ref{tab:bug-reports}.  According to the various fields in Figure \ref{front_end} the reporter would first fill in their (i) \textit{name} (Field 1), (ii) \textit{device} (Field 2), (iii) \textit{screen orientation} (Field 3), (iv) a \textit{bug report title} (Field 4), and (v) a \textit{brief description of the bug} (Field 5).

\subsection{Auto-completing Bug Reproduction Steps}
\label{auto-completion}

  To facilitate the reporter in entering reproduction steps, we model each step in the reproduction process as an {\tt \{action, component\}} tuple corresponding to the action the reporter wants to describe at each step, (e.g., tap, long-tap, swipe, type) and the component in the app GUI with which they interacted (e.g.,``Name" textview, ``OK" button, ``Days" spinner).  Since reporters are generally aware of the actions and GUI elements they interact with, it follows that this is an intuitive manner for them to construct reproduction steps.  FUSION allocates auto-completion suggestions to drop down lists based on a decision tree taking into account a reporter's position in the app execution beginning from a cold-start of the app.
     
\begin{figure}[tb]
\centering
\includegraphics[width=\linewidth]{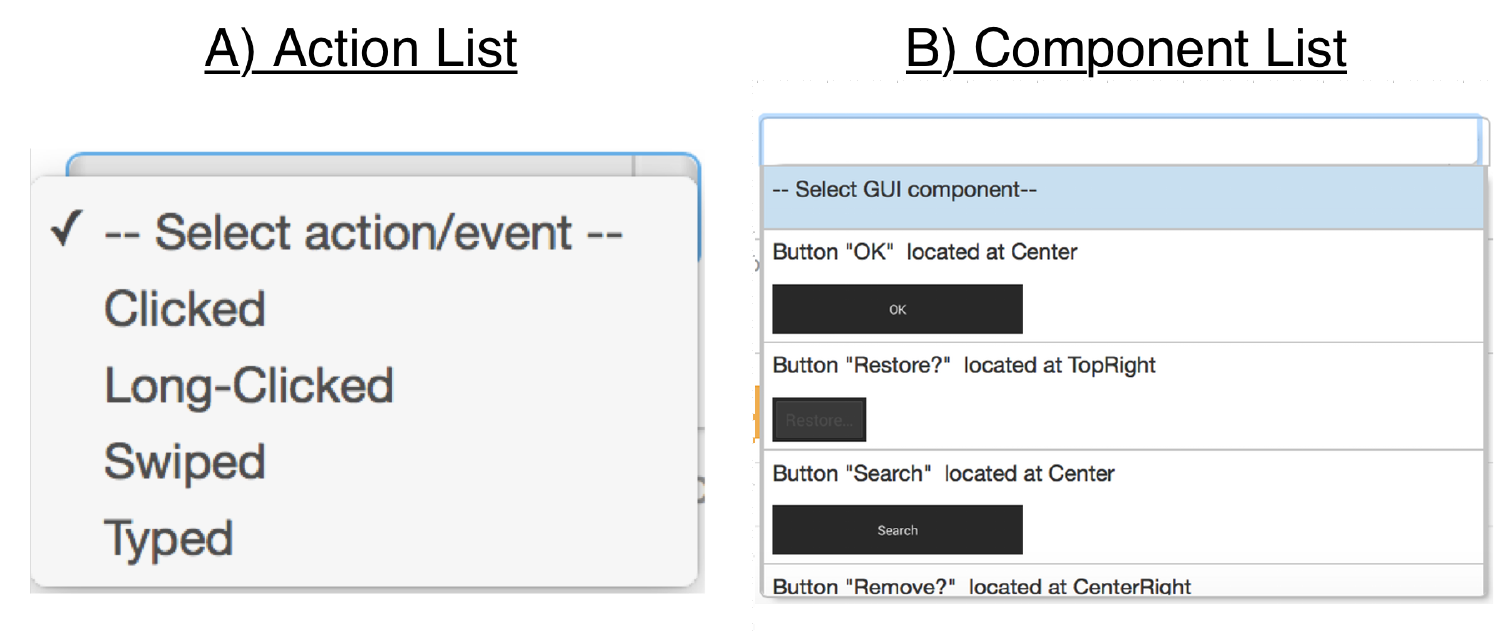}
\vspace{-0.7cm}
\caption{\textbf{Auto-Completion Dropdown Menus:} This figure shows examples of the auto-completion dropdown menus that present reporters with possible choices during course of creating a report.}
\label{autocomplete_box}
\end{figure}

     The first drop down list (see Figure \ref{autocomplete_box}-A) corresponds to the possible actions a user can perform at a given point in app execution.  In our example with the Document Viewer bug, let's say the reporter selects \textit{click} as the first action in the sequence of steps as shown in Figure \ref{autocomplete_box}-A. The possible actions considered in FUSION  are \textit{click(tap), long-click(long-touch), type}, and \textit{swipe}.  The \textit{type} action corresponds to a user entering information from the device keyboard.  When the reporter selects the \textit{type} option, we also present them with a text box to collect the information they typed in the Android app.

	The second dropdown list (see Figure \ref{autocomplete_box}-B) corresponds to the component associated to the action in the step. FUSION presents the following information, which can also be seen in Figure \ref{autocomplete_box}: (i) \textit{Component Type}: this is the type of component that is being operated upon, e.g., button, spinner, checkbox,  (ii) \textit{Component Text}: the text associated with or located on the component, (iii) \textit{Relative Location}: the relative location of the component on the screen according to the parameters in Figure \ref{example_screenshot}, and (iv) \textit{Component Image}: an in-situ (i.e., embedded in the dropdown list) image of the instance of the component.  The relative location is displayed here to make it easier for reporters to reason about the on-screen location, rather than reasoning about pixel values.  In our running example, FUSION will populate the component dropdown list with all of the clickable components in the Main Activity since this is the first step and the selected action was \textit{click}.  The user would then select the component they acted upon, in this case, the first option in the list: the ``OK" button located at the center of the screen (see Figure \ref{autocomplete_box}-B).
    
    One potential issue with component selection from the auto-complete drop-down list is that there may be duplicate components on the same screen in an app.  FUSION solves this problem in two ways. \textit{First}, it differentiates each duplicate component in the list through specifying text ``Option \#''.  \textit{Second} FUSION attempts to confirm the component entered by the reporter at each step by fetching screen-shots from the FUSION database representing the entire device screen.  Each of these screen-shots highlights the representative GUI component as shown in Fig. \ref{example_screenshot}. To complete the step entry the reporter simply selects the screen-shot corresponding to both the app state and the GUI component acted upon.  In our running example the reporter would select the full augmented screenshot corresponding to the component they selected from the dropdown list. In our case an illustrative portion of the screenshot for the ``OK" button is shown in Figure \ref{example_screenshot}.
    
     After the reporter makes selections from the drop-down lists, they have an opportunity to enter additional information for each step (e.g., a button had an unexpected behavior) in a natural language text entry field.  For instance in our running example, the reporter might indicate that after pressing the ``OK" button the pop-up window took longer than expected to disappear.  

\subsection{Report Generator Auto-Completion Engine}
\label{auto-engine}

The \emph{Auto-Completion Engine} of the web-based report generator (Figure \ref{Design}-\circled{4}) uses the information collected up-front during the \textit{Analyze Phase}. When FUSION suggests completions for the drop-down menus it queries the database for the corresponding state of the app event flow, and suggests information based on the past steps that the reporter has entered.  Since we always assume a ``cold'' application start, the \emph{Auto-Completion Engine} starts the reproduction steps entry process from the app's main Activity.  We then track the reporter's progress through the app using predictive measures based on past steps.  
 
\begin{figure}[tb]
\centering
\includegraphics[width=\linewidth]{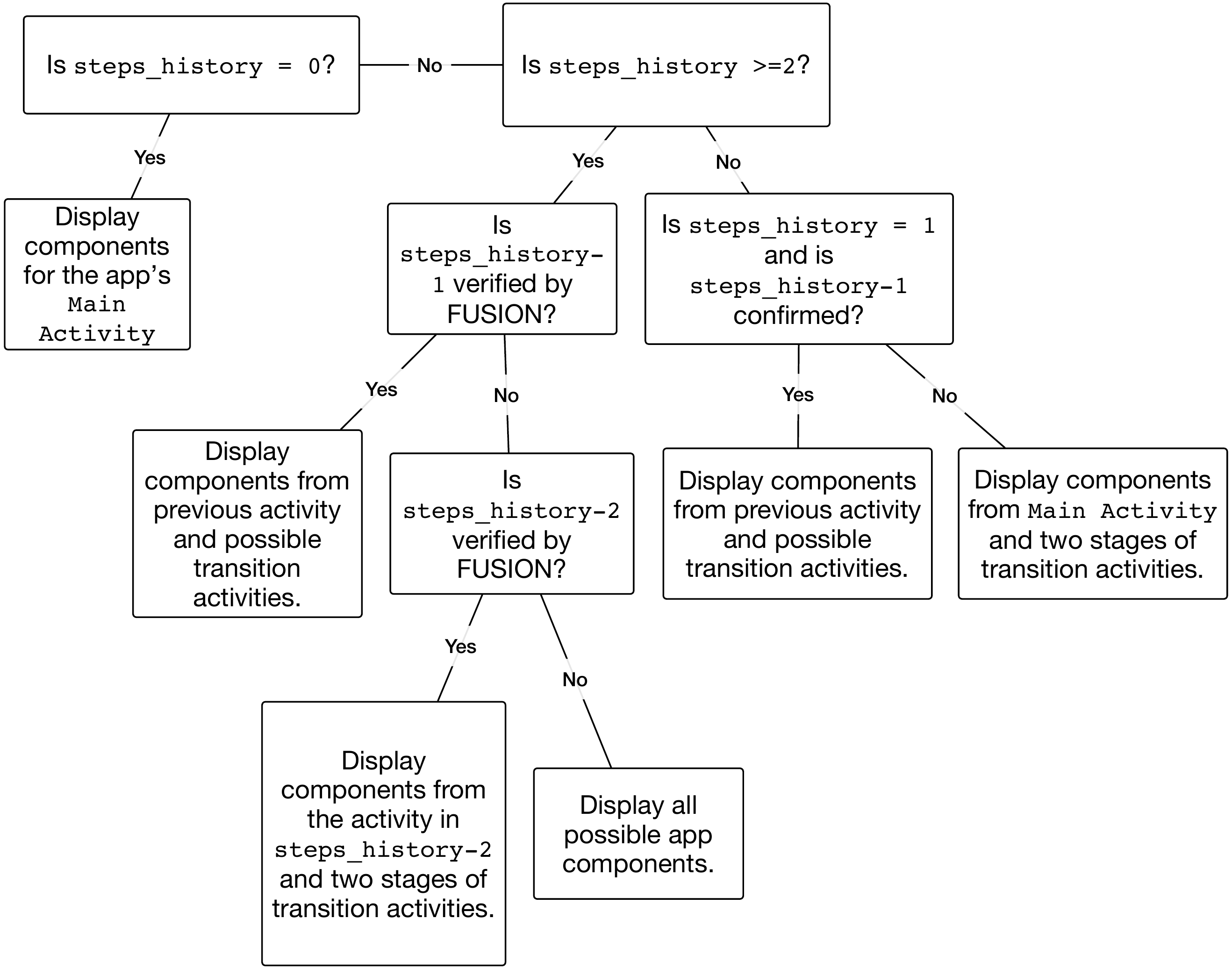}
\vspace{-0.6cm}
\caption{\textbf{Decision Tree Utilized by Auto-Completion Engine:}  This figure outlines the decision tree utilized by FUSION's autocompletion engine which helps to predict the possible components that a user can interact with at a specific place in the event flow of an application.}
\label{decision_tree}
\end{figure}

 	The \emph{Auto-Completion Engine} operates on application steps using several different pieces of information as input.  It models the reporter's reproduction steps as an ordered stream of steps $S$ where each individual step $s_i$ may be either empty or full.  Each step can be modeled as a five-tuple consisting of \textit{\{step\_num, action, comp\_name, activity, history\}}.  The \textit{action} is the gesture provided by the reporter in the first drop-down menu.  The \textit{component\_name} is the individual component name as reported by the {\tt UIautomator} interface during the Engine phase.  The \textit{activity} is the Android screen the component is found on.  The \textit{history} is the history of steps preceding the current step.  The auto-completion engine predicts the suggestion information using decision tree logic which can be seen in Figure \ref{decision_tree}.  
	
	FUSION presents components to the reporter at the granularity of activities or application screens.  To summarize the suggestion process, FUSION looks back through the history of the past few steps and looks for possible transitions from the previous steps to future steps depending on the components interacted with.  If FUSION was unable to capture the last few steps from the reporter due to the incomplete application execution model mentioned earlier, then FUSION presents the possibilities from all known screens of the application. In our running example, let's consider the reporter moving on to report the second reproduction step.  In this case, FUSION would query the history to find the previous activity the ``OK" button was located within, and then present component suggestions from that activity, in the case that the user stayed in the same activity; and the components from possible transition activities, in the case the user transitioned to a different activity.

\subsection{Handling FUSION's Application Model Gaps}
\label{gap_handling}

\begin{figure}[h]
\centering
\includegraphics[width=\linewidth]{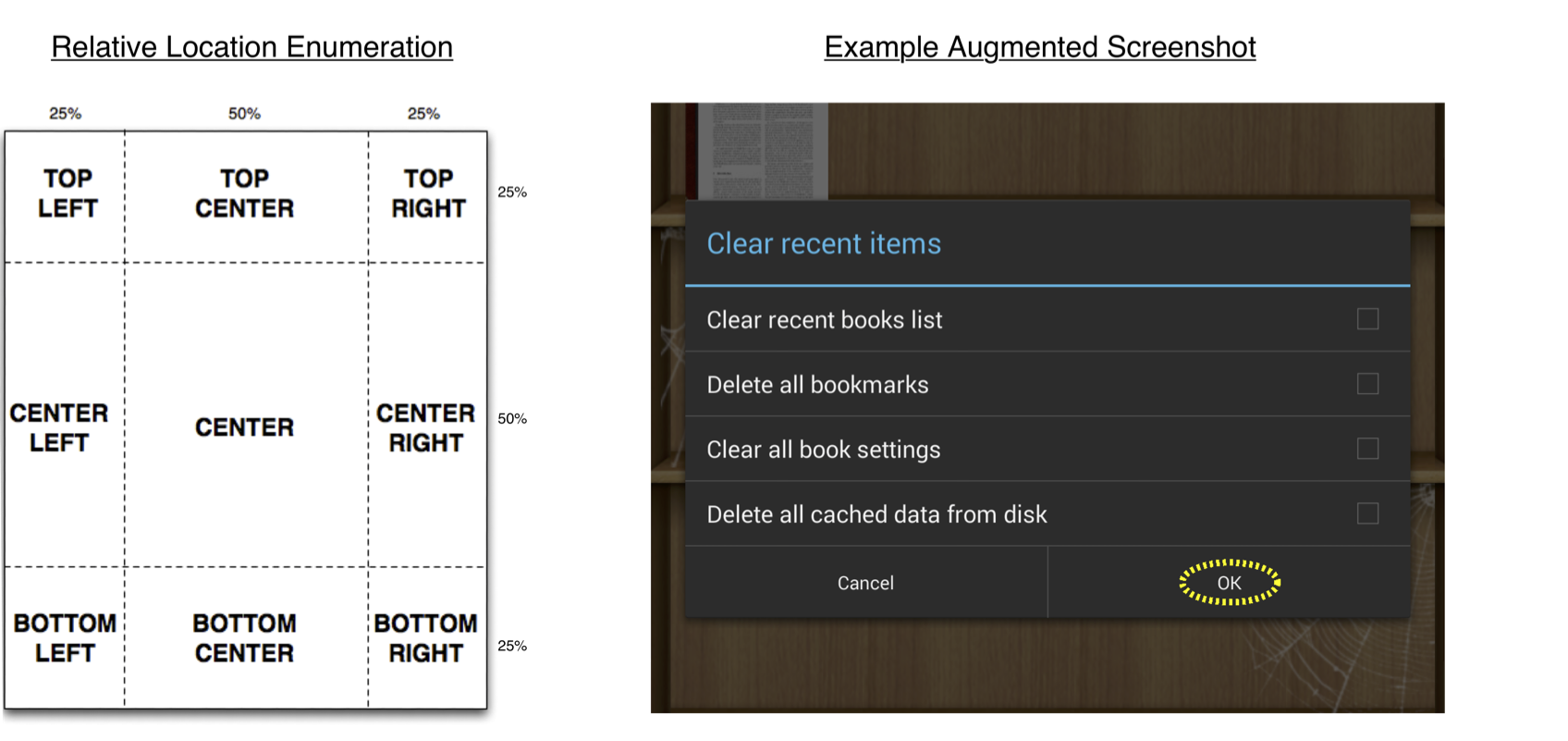}
\vspace{-0.8cm}
\caption{\textbf{Relative Location Enumeration and Example Augmented Screenshot:}  This figure shows FUSION's enumeration for the relative location of GUI-components on the screen and an example of an augmented full screenshot.}
\label{example_screenshot}
\end{figure}

 Because DFS-based exploration is not exhaustive \cite{Nguyen:TSE14}, there may be gaps in FUSION's database of possible app screens (e.g., a dynamically generated component that triggers an activity transition was not acted upon). Due to this, a reporter may not find the appropriate suggestion in the drop-down list.  To handle these cases gracefully, we allow the reporter to select a special option when they cannot find the component they interacted with in the auto-complete drop-down list. In our running example, let's say the reporter wishes to indicate that he clicked the button labeled ``Open Document", but the option is not available in the auto-complete component drop-down list.  In this case the user would select the ``Not in this list...'' option and manually fill in  (i) The type of the component (to limit confusion, we present this option as a drop-down box auto-completed with only the GUI-component types that exist in the application, as extracted by the \emph{Primer}, in our case the user would choose "Button"), (ii) any text associated with the GUI-component (in this case ``Open Document", and (iii) the relative location of the GUI-component as denoted in Figure \ref{example_screenshot} (in this case ``Top Center").
    
\subsection{Report Structure}
\label{reports}

\begin{figure}[p]
\centering
\includegraphics[width=\linewidth]{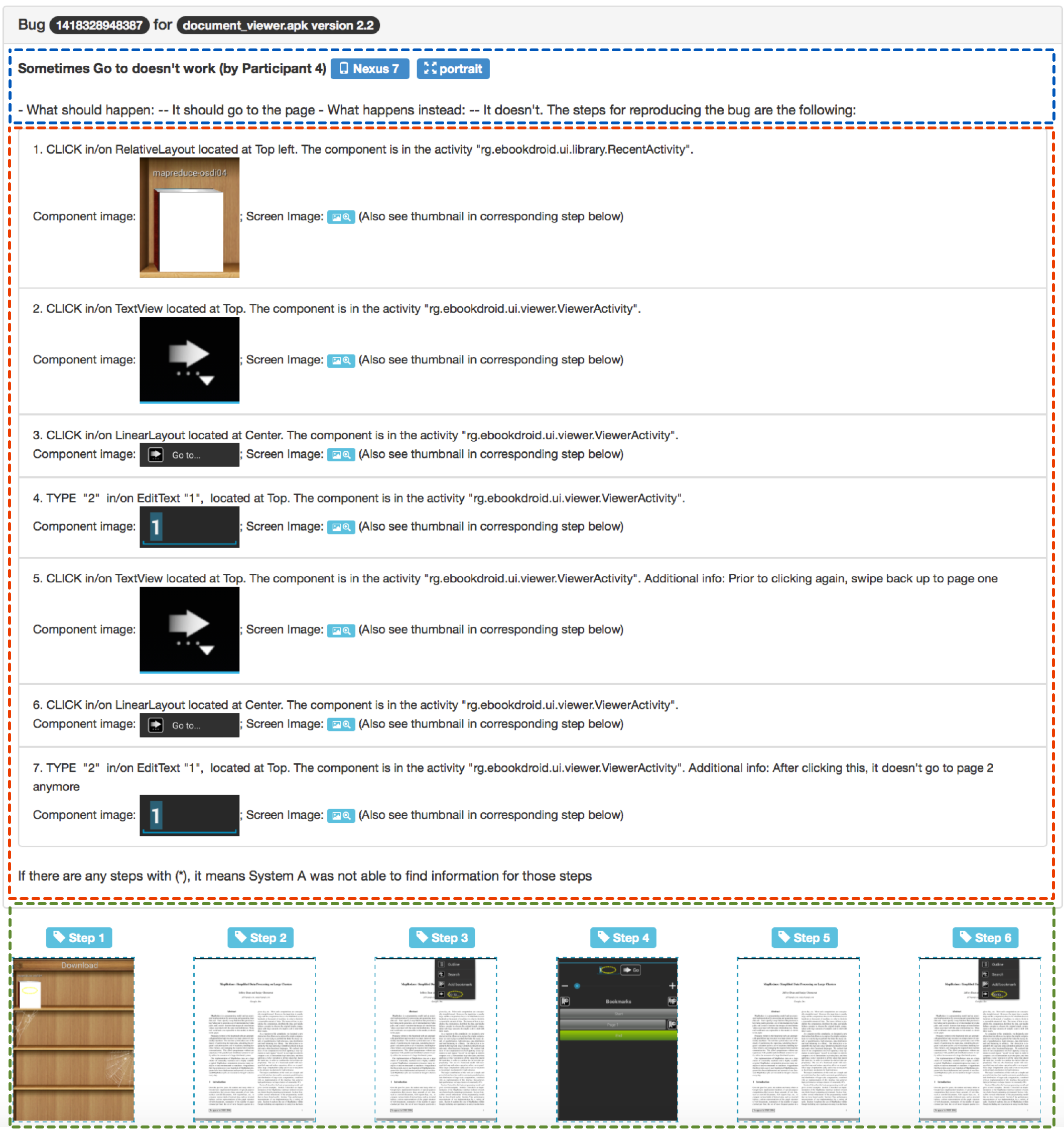}
\vspace{-0.4cm}
\caption{\textbf{Example FUSION Bug Report:}  This report shows the three major categories of information contained within FUSION bug reports 1) Contextual information and natural language description of the bug.  2) A detailed set of reproduction steps including GUI-component specific screenshots. 3) A list of fullscreen screenshots with the GUI-component acted upon in each step highlighted.}
\label{report}
\end{figure}

 The \emph{Auto Completion Engine} saves each step to the database as reporters complete bug reports.  Once a reporter finishes filling out the steps and completes the data entry process, a screen containing the final report, with an automatically assigned unique ID, is presented to the reporter, and saved to the database for a developer to view later (see Figure \ref{report} for an example report from Document Viewer). The Report presents information to developers in three major sections. First, preliminary information including the report title, device, and short description (shown in Figure \ref{report} in blue).  Second, a list of the Steps with the following information regarding each step is dispalyed (highlighted in blue in Figure \ref{report}): (i) The Action for each step, (ii) the type of a component, (iii) the relative location of the component, (iv) the Activity class where the component is instantiated in the source code, and (v) the component specific screenshot.  Third, a list of full screen-shots corresponding to each step is presented at the bottom of the page so the developer can trace the steps through each application screen (this section is highlighted in green in Figure \ref{report}). 
    
\chapter{Design of Empirical Studies}
\label{study_design}

The two major design goals behind FUSION are: \textbf{1)} \textit{facilitate and encourage reporters to submit useful bug reports for Android applications.}
\textbf{2)}  \textit{provide developers with more actionable information regarding the bugs contained within these reports.} In order to measure FUSION's effectiveness at achieving these goals, we have designed two comprehensive empirical studies which evaluated two major aspects of our approach: 1) \textit{the user experience of reporters using FUSION}, and 2) \textit{the quality of the bug reports produced by the system} . To this end, we investigated the following research questions (RQs):
  
  \begin{itemize}
\item \textbf{RQ$_1$}: \textit{What types of information fields do developers/testers consider important when reporting and reproducing bugs in Android?}
\item \textbf{RQ$_2$}: \textit{Is FUSION easier to use for reporting and reproducing bugs than traditional bug tracking systems?}
\item \textbf{RQ$_3$}: \textit{Do bug reports generated with FUSION allow for faster bug reproduction compared to reports submitted using traditional bug tracking systems?}
\item \textbf{RQ$_4$}: \textit{Do developers/testers using FUSION reproduce more bugs compared to traditional bug tracking systems?} 
\end{itemize}

The empirical studies used to evaluate these research questions model two maintenance activities involving reporting and reproducing real bugs in open source apps. In the following sections we will describe the context of the two studies (i.e., Android apps and bug reports), and the methodology of each study.

\section{Study Context: Bug Reports Used in the Studies}
\label{bug_reports}

In order to properly evaluate FUSION for creating and reproducing reports from real world bugs, we manually selected bug reports from Android Open Source apps hosted on the F-Droid \cite{fdroid} repository. We crawled the links of the issue tracking systems of the apps, and then manually inspected the bug reports for each project where F-droid has a linked issue tracker.  The criteria for selecting the bug reports were the following: 1) bugs that are reproducible given the technical constraints of our FUSION implementation; 2) bugs of varying complexity, requiring at least three steps of user interaction in order to be manifested; and 3) bugs that are reproducible on the Nexus 7 tablets utilized for the user study. Details of these bug reports can be found in Table \ref{tab:bug-reports} and links can be found in our online appendix \cite{appendix}.  
	
	FUSION targets bug reports that can be described in terms of GUI events and are not context dependent.  For instance, some bugs are triggered when changing the orientation of the device, or are context dependent (i.e., the bug depends on the network signal quality, GPS location, etc.). We do not claim that our FUSION approach works for all types of Android bugs, but rather acknowledge and give examples of the current limitations in Chapter \ref{limitations}.  However, even in cases where FUSION may not be able to capture the exact cause of the bug, the steps to reproduction, along with additional information added to the last step may aid in reproducing and fixing various types of mobile bugs. Application activity coverage statistics can be found in Table \ref{tab:bug-reports}.  We present activity coverage information in this table to give context describing the extent to which FUSION's dynamic analysis \textit{Engine} was able to explore the app.  Due to the nature of our DFS app traversal, most of the components within an activity are explored.

\begin{table*}[p]
\centering
\begin{small}
\caption{\textbf{Summary of the bug reports used for the empirical studies:}  GDE = Gui Display Error, C = Crash, DIC = Data Input/Calculation Error, NE = Navigation Error;  (Links to the original bug reports can be found at our online appendix)}
\label{tab:bug-reports}
\begin{tabular}{ | l | p{20pt} | p{110pt} | p{30pt} | p{40pt} | p{60pt} |}
\hline
\textbf{App (Bug Index)} & \textbf{Bug ID} & \textbf{Description} & \textbf{Min \# of Steps} & \textbf{Bug Type} & \textbf{DFS Activity Coverage}\\ \hline
1) A Time Tracker & 24 & Dialog box is displayed three times in error. & 3 & GDE & 1/5\\ \hline
2) Aarddict & 106 & Scroll Position of previous pages is incorrect. & 4-5
 & GDE& 3/6\\ \hline
3) ACV  & 11 & App Crashes when long pressing on sdcard folder. & 5 & C & 3/11 \\ \hline
4) Car report & 43 & Wrong information is displayed if two of the same values are entered subsequently & 10 & DIC & 5/6\\ \hline
5) Document Viewer & 48 & Go To Page \# number requires two entries before it works & 4 & NE  & 4/8\\ \hline
6) DroidWeight & 38 & Weight graph has incorrectly displayed digits & 7 & GDE  & 3/8\\ \hline
7) Eshotroid & 2 & Bus time page never loads. & 10 & GDE/NE & 6/6 \\ \hline
8) GnuCash & 256 & Selecting from autocomplete suggestion doesn't allow modification of value & 10 & DIC & 3/4\\ \hline
9) GnuCash & 247 & Cannot change a previously entered withdrawal to a deposit. & 10 & DIC  & 3/4\\ \hline 
10) Mileage & 31 & Comment Not Displayed. & 5 & GDE/DIC & 2/27 \\ \hline 
11) NetMBuddy & 3 & Some YouTube videos do not play. & 4 & GDE/NE & 5/13\\ \hline 
12) Notepad & 23 & Crash on trying to send note. & 6 & C  & 4/7\\ \hline 
13) OI Notepad & 187 & Encrypted notes are sorted in random when they should be ordered alphabetically & 10 & GDE/DIC & 3/9\\ \hline 
14) Olam & 2  & App Crashes when searching for word with apostrophe or just a "space" character & 3 & C & 1/1\\ \hline 
15) QuickDic & 85 & Enter key does not hide keyboard & 5 & GDE & 3/6\\ \hline
\end{tabular}
\end{small}
\vspace{-0.5cm}
\end{table*}

\section{Study 1: Reporting Bugs with FUSION}
\label{study1}

The \textit{\textbf{goal}} of the first study is to assess whether FUSION's features are useful when reporting bugs for Android apps, which aims to address \textbf{RQ$_1$ \& RQ$_2$}. In particular, we want to identify if the auto-completion steps and in-situ screenshot features are useful when reporting bugs. For this, we recruited eight students (four undergraduate or \textit{non-experts} and four graduate or \textit{experts}) at the College of William and Mary to construct bug reports using FUSION and the Google Code Issue Tracker (GCIT) --- as a representative of traditional bug tracking systems--- for the real world bugs from the reports shown in Table \ref{tab:bug-reports}. We chose the Google Code Issue tracker as our comparison benchmark as it represents a general standard for current issue tracking systems in terms of features and is widely used in many open source software projects.  The four graduate participants had extensive programming backgrounds. Four participants constructed a bug report for each of the 15 bugs in Table \ref{tab:bug-reports} using FUSION prototype, and four participants reported bugs using the Google Code Issue Tracker Interface.  The participants were distributed to the systems to have non-experts and programmers evaluating both systems. The Design Matrix for this phase of the study can be seen in Table \ref{tab:s1-design_matrix}.  In total the participants constructed 60 bug reports using FUSION and 60 using GCIT.  Participants used a Nexus 7 tablet with Android 4.4.3 KitKat installed to reproduce the bugs.  

\begin{table}[t]
\caption{\textbf{Study 1 Participant Design Matrix:} This table shows the indicies of bug reports assigned to participants during Study 1.  The bugs corresponding to the index numbers can be found in Table \ref{tab:bug-reports} }
\vspace{-0.3cm}
\small
\begin{center}
\begin{tabular}{ | l | l | l | l | }
\hline
	\textbf{Phase 1: Creation} & \textbf{Participant} & \textbf{Report Type} & \textbf{Bug Numbers (Index)}\\ \hline
	Experienced Users & 1 & FUSION(E)1 & 1-15 \\ \hline
	 & 2 & FUSION(E)2 & 1-15 \\ \hline
	 & 3 & Google Code((E)1 & 1-15 \\ \hline
	 & 4 & Google Code (E)2 & 1-15 \\ \hline
	Non-experienced Users & 5 & FUSION (I)1 & 1-15 \\ \hline
	 & 6 & FUSION (I)2 & 1-15 \\ \hline
	 & 7 & Google Code (I)1 & 1-15 \\ \hline
	 & 8 & Google Code (I)2 & 1-15 \\ \hline
\end{tabular}
\end{center}
\label{tab:s1-design_matrix}
\vspace{-0.3cm}
\end{table}%

	 One challenge in conducting this first study is illustrating the bug to the participants without introducing bias from the original bug report.  In other words, we wanted the user not to create a bug report from a bug report, but rather create a bug report through experiencing the bug naturally.  To accomplish this, we created a short video of the steps to reproduce the bug. So as not to influence the complexity of the bugs, we recorded the videos using the fewest possible number of user steps to manifest the bug in question.  After the users experienced the bug through the video, they were asked to confirm it by reproducing the bug on the loaned Nexus 7 tablet.  After the users manifested the bug they were asked to construct the bug report for the corresponding system to which they were assigned.  During the reports collection, the names of the bug reporting systems were anonymized to ``System A" for FUSION and ``System B" for GCIT. The users were provided with a short tutorial regarding how to enter bugs for each system, so as not to introduce bias towards any reporting system. 
	 	 
	 In addition to the bug reports, we collected the amount of time it took each participant to fill out each bug report, as well as responses to a set of questions after filling out all of the bug reports for the system.  The questions were focused on three different aspects: 1) user preferences, 2) user experience and 3) demographic background. The preferences-related questions  were formulated based on the user experience honeycomb originally developed by Peter Moville \cite{Morville:04}, The preferences-related questions are listed in Table  \ref{tab:up1-questions}.  The usability was evaluated by using statements based on the SUS usability scale by John Brooke \cite{Brooke:96}. These statements are listed in Table \ref{tab:ux1-questions}. The questions were used to evaluate the user experience with the systems and  were presented to participants replacing the token \texttt{(system)} with the anonymized name of the system they were evaluating (i.e., System A or System B).  The full instructions that were used during this user study can be found in Appendix \ref{appen:user_instrucs}.

\begin{table}[t]
\caption{\textbf{Study 1 User Preference Questions:} Questions used during Study 1 to evaluate User Preferences regarding FUSION.}
\vspace{-0.3cm}
\small
\begin{center}
\begin{tabular}{ | l | l | }
\hline
	\textbf{Question ID} & \textbf{Question} \\ \hline
	S1UP1 & What fields in the form did you find useful when reporting the bug? \\ \hline
	S1UP2 (FUSION ONLY) & Were the component suggestions accurate? \\ \hline
	S1UP3 (FUSION ONLY) & Were the screenshot suggestions accurate? \\ \hline
	S1UP4 & What information if any were you not able to report? \\ \hline
	S1UP4 & What elements do you like most from the system? \\ \hline
	S1UP5 & What elements do you like least in the system? \\ \hline
	S1UP6 & Please give any additional feedback about the bug reporting system? \\ \hline
\end{tabular}
\end{center}
\label{tab:up1-questions}
\vspace{-0.3cm}
\end{table}%

\begin{table}[t!]
\caption{\textbf{Study 1 User Experience Questions:} Questions used during Study 1 to evaluate the User Experience of FUSION.}
\vspace{-0.3cm}
\small
\begin{center}
\begin{tabular}{ | l | l | }
\hline
	\textbf{Question ID} & \textbf{Question} \\ \hline
	S1UX1 & I think that I would like to use (system) frequently. \\ \hline
	S1UX2 & I found (system) very cumbersome to use. \\ \hline
	S1UX3 & I found the various functions in (system) were well integrated. \\ \hline
	S1UX4 & I thought (system) was easy to use. \\ \hline
	S1UX5 & I found (system) unnecessarily complex. \\ \hline
	S1UX6 & I thought (system) was really useful for reporting a bug. \\ \hline
\end{tabular}
\end{center}
\label{tab:ux1-questions}
\vspace{-0.3cm}
\end{table}%
 
 \begin{table}[t!]
\caption{\textbf{Participant Programming Experience Questions} Questions used to evaluate the relative programming experience of participants in both empirical studies.}
\vspace{-0.3cm}
\small
\begin{center}
\begin{tabular}{ | l | p{370pt} | }
\hline
	\textbf{Question ID} & \textbf{Question} \\ \hline
	PX1 & On a scale of 1 to 10 how do you estimate your programming experience? (1: very inexperienced 10: very experienced)  \\ \hline
	PX2 & On a scale of 1 to 10 how experienced are you with Android programming paradigms? (1: very inexperienced 10: very experienced) \\ \hline
	PX3 & For how many years have you been programming? \\ \hline
	PX4 & For how many years have you been studying computer science? \\ \hline
	PX5 &  How many courses (roughly) have you taken in which you had to write source code? \\ \hline
\end{tabular}
\end{center}
\label{tab:ux2-questions}
\vspace{-0.3cm}
\end{table}%

   The questions for user preferences (UP questions in Table \ref{tab:ux1-questions}) were free form text entry fields, the user experience questions (UX Questions in Table \ref{tab:ux1-questions}) and programming experience was scored by the participant on a Likert scale (1 representing a strong disagreement and 5 representing strong agreement). Background information questions are based on the programming experience questionnaire developed by Feigenspan et al \cite{Feigenspan:ICPC12}. For the analysis of the open questions, one of the authors analyzed and categorized the answers manually.  The results of this study, and their applicability to the research questions are discussed in Chapter \ref{study_results}.

\section{Study 2: Reproducibility of Bug Reports}
\label{study2}

Whereas Study 1 analyzes FUSION from the viewpoint of a reporter, Study 2 is centered on developers and the activity of reproducing bugs, which corresponds specifically to \textbf{RQ$_2$}-\textbf{RQ$_4$}. However, the information collected during this study has bearing on all research questions, and the relevant information gleaned from the study and its applicability to the research questions is described in Chapter \ref{study_results} Therefore, the \emph{goal} of Study 2 is to evaluate the ability of our proposed approach to improve the reproducibility of bug reports. In particular, we evaluated the following aspects in FUSION and traditional issue trackers: 1) usability when using the bug tracking systems' GUIs for reading bug reports, 2) time required to reproduce reals bugs by using the bug reports, and 3) number of bugs that were successfully reproduced.  The reports generated during Study 1, using FUSION and GCIT, in addition to the original bug reports (Table~\ref{tab:bug-reports}) were evaluated by a new set of participants by attempting to reproduce the bugs on physical devices.
 
\begin{table}[t]
\caption{\textbf{Study 2 User Preference Questions:} Questions used during Study 1 to evaluate User Preferences regarding FUSION.}
\vspace{-0.3cm}
\small
\begin{center}
\begin{tabular}{ | l | p{300pt} | }
\hline
	\textbf{Question ID} & \textbf{Question} \\ \hline
	S2UP1 & What information from this type of Bug Report did you find useful for reproducing the bug? \\ \hline
	S2UP2 & What other information if any would you like to see in this type of bug report? \\ \hline
	S2UP3 & What elements did you like the most from this type of bug report? \\ \hline
	S2UP4 & What information did you like least from this type of bug report? \\ \hline
\end{tabular}

\end{center}
\label{tab:up2-questions}
\vspace{-0.3cm}
\end{table}%

\begin{table}[t!]
\caption{\textbf{Study 2 User Experience Questions:} Questions used during Study 1 to evaluate the User Experience of FUSION.}
\vspace{-0.3cm}
\small
\begin{center}
\begin{tabular}{ | l | l | }
\hline
	\textbf{Question ID} & \textbf{Question} \\ \hline
	S2UX1 & I think that I would like to use this type of bug report frequently. \\ \hline
	S2UX2 & I found this type of bug report unnessecarily complex. \\ \hline
	S2UX3 & I thought this type of bug report was easy to read/understand. \\ \hline
	S2UX4 & I found this type of bug report very cumbersome to read. \\ \hline
	S2UX4 & I thought the bug report was really useful for reproducing the bug. \\ \hline
\end{tabular}

\end{center}
\label{tab:ux2-questions}
\vspace{-0.3cm}
\end{table}%

For the evaluation we enlisted 20 new participants, none of which participated in the first study. The participants were graduate students from the Computer Science Department at College of William and Mary, all of whom are familiar with the Android platform. All participants were compensated \$15 USD for their efforts.  Each user evaluated 15 bug reports, six from FUSION, six from GCIT, and three original.  135 reports were evaluated (120 from Study 1 plus the 15 original bug reports), and were distributed to the 20 participants in such a way that each bug report was evaluated by two different participants (the full design matrix can be found in our online appendix \cite{appendix}).  Each participant evaluated only one version of a bug report for a bug, since due to the learning effect, after a user reproduces a bug once, they will be capable of reproducing it easily in subsequent attempts with other bug reports. To clarify, if a participant $p$ analyzed a bug report typed in system $A$ for bug $x$, no other bug report for bug $x$ was assigned to $p$. The full design matrix for this study can be seen in Table \ref{tab:study2_design_matrix}.  

\begin{table}[t]
\caption{\textbf{Study 2 Participant Design Matrix}: This table shows the indicies of bug reports assigned to participants during Study 2.  The bugs corresponding to the index numbers can be found in Table \ref{tab:bug-reports}}
\vspace{-0.3cm}
\scriptsize
\begin{center}
\begin{tabular}{ | p{60pt} | l | p{50pt} | p{50pt} | l | p{50pt} |}
\hline
	 \textbf{Inexperienced Participants} & \textbf{Report Type} & \textbf{Bug \#'s (Index)} & \textbf{Experienced Participants} & \textbf{Report Type} & \textbf{Bug \#'s (Index)} \\ \hline
	 1 & Original  & 1-3  & 11 & Original & 1-3  \\ \hline
	   & Google Code (E) 1 & 4-6  &  & Google Code (E) 2 & 4-6 \\ \hline
	   & Google Code (I) 1 & 7-9  &  & Google Code (I) 2 & 7-9\\ \hline
	   & FUSION(E) 1 & 10-12 &  & FUSION(E) 2 & 10-12\\ \hline
	   & FUSION (I) 1 & 13-15 &  & FUSION (I) 2 & 13-15  \\ \hline
	  2 & Google Code (E) 1 & 1-3 & 12 & Google Code (E) 2 & 1-3 \\ \hline
	   & Google Code (I) 1 & 4-6 &  & Google Code (I) 2 & 4-6 \\ \hline
	   & FUSION(E) 1 & 7-9 &  & FUSION(E) 2 & 7-9 \\ \hline
	   & FUSION (I) & 10-12  &  & FUSION (I) 2 & 10-12\\ \hline
	   & Original & 13-15  &  & Original & 13-15 \\ \hline
	  3 & Google Code (I) 1 & 1-3 & 13 & Google Code (I) 2 & 1-3\\ \hline
	   & FUSION(E) 1 & 4-6 &  & FUSION(E) 2 & 4-6 \\ \hline
	   & FUSION (I) 1 & 7-9 &  & FUSION (I) 2 & 7-9\\ \hline
	   & Original & 10-12 &  & Original & 10-12\\ \hline
	   & Google Code (E) 1 & 13-15  &  & Google Code (E) 2 & 13-15\\ \hline
	  4 & FUSION(E) 1 & 1-3  & 14 & FUSION(E) 2 & 1-3 \\ \hline
	   & FUSION (I) 1 & 4-6 &  & FUSION (I) 2 & 4-6  \\ \hline
	   & Original & 7-9  &  & Original & 7-9\\ \hline
	   & Google Code (E) 1 & 10-12 &  & Google Code (E) 2 & 10-12 \\ \hline
	   & Google Code (I) 1 & 13-15 &  & Google Code (I) 2 & 13-15\\ \hline
	  5 & FUSION (I) 1 & 1-3 & 15 & FUSION (I) 2 & 1-3 \\ \hline
	   & Original & 4-6  &  & Original & 4-6\\ \hline
	   & Google Code (E) 1 & 7-9  &  & Google Code (E) 2 & 7-9\\ \hline
	   & Google Code (I) 1 & 10-12 &  & Google Code (I) 2 & 10-12\\ \hline
	   & FUSION(E) 1 & 13-15 &  & FUSION(E) 2 & 13-15\\ \hline
	  6 & Original & 1-3 & 16 & Original & 1-3\\ \hline
	   & Google Code (E) 1 & 4-6 &  & Google Code (E) 2 & 4-6  \\ \hline
	   & Google Code (I) 1 & 7-9 &  & Google Code (I) 2 & 7-9\\ \hline
	   & FUSION(E) 1 & 10-12 &  & FUSION(E) 2 & 10-12\\ \hline
	   & FUSION (I) 1 & 13-15  &  & FUSION (I) 2 & 13-15 \\ \hline
	  7 & Google Code (E) 1 & 1-3 & 17 & Google Code (E) 2 & 1-3 \\ \hline
	   & Google Code (I) 1 & 4-6  &  & Google Code (I) 2 & 4-6\\ \hline
	   & FUSION(E) 1 & 7-9 &  & FUSION(E) 2 & 7-9 \\ \hline
	   & FUSION (I) 1 & 10-12 &  & FUSION (I) 2 & 10-12 \\ \hline
	   & Original & 13-15 &  & Original & 13-15\\ \hline
	  8 & Google Code (I) 1 & 1-3 & 18 & Google Code (I) 2 & 1-3\\ \hline
	   & FUSION(E) 1 & 4-6  &  & FUSION(E) 2 & 4-6 \\ \hline
	   & FUSION (I) 1 & 7-9 &  & FUSION (I) 2 & 7-9\\ \hline
	   & Original & 10-12 &  & Original & 10-12\\ \hline
	   & Google Code (E) 1 & 13-15 &  & Google Code (E) 2  & 13-15\\ \hline
	  9 & FUSION(E) 1 & 1-3 & 19 & FUSION(E) 2 & 1-3\\ \hline
	   & FUSION (I) 1 & 4-6 &  & FUSION (I) 2 & 4-6\\ \hline
	   & Original & 7-9 &  & Original & 7-9\\ \hline
	   & Google Code (E) 1 & 10-12 &  & Google Code (E) 2  & 10-12\\ \hline
	   & Google Code (I) 1 & 13-15&  & Google Code (I) 2 & 13-15 \\ \hline
	  10 & FUSION (I) 1 & 1-3 & 20 & FUSION (I) 2 & 1-3\\ \hline
	   & Original & 4-6 &  & Original & 4-6\\ \hline
	   & Google Code (E) 1 & 7-9  &  & Google Code (E) 2 & 7-9\\ \hline
	   & Google Code (I) 1 & 10-12 &  & Google Code (I) 2 & 10-12\\ \hline
	   & FUSION(E) 1 & 13-15 &  &  FUSION (E) 2 & 13-15\\ \hline
	 \end{tabular}
\end{center}
\label{tab:study2_design_matrix}
\vspace{-0.3cm}
\end{table}%
	
During the study, the participants were sent links corresponding to the reports for which they were tasked with reproducing the bug.  Each participant was loaned a Nexus 7 tablet with Android 4.4.3 KitKat installed; the apps were preinstalled in the devices.  For each bug report, the users attempted to recreate the bug on the tablet device using only the information contained within the report.  The users timed themselves in the reproduction for each bug, with a ten minute time limit.  If a participant was not able to reproduce a bug after ten minutes, that bug was marked as \textit{not-reproduced}.  A proctor monitored the study to judge whether participants successfully reproduced a given bug.  After the users attempted to reproduce all 15 bugs assigned to them, they were asked to fill out an anonymous online questionnaire for each type of the bug report they utilized, with the UX and UP questions in Tables \ref{tab:up2-questions} and \ref{tab:ux2-questions}.  The full set of user instructions that utilized during this study can be found in Appendix \ref{appen:user_instrucs}.
As for the analysis,  we used descriptive statistics to analyze the responses for the UX statements, the time for reproducing the bugs, and the number of successful reproductions. Results for this study are presented in Chapter \ref{study_results}.

\chapter{Empirical Study Results}
\label{study_results}

\section{Study 1 (Bug Report Creation) Results}
\label{results1}

In this section we present the qualitative and quantitative results for \textit{Empirical Study 1}. We begin with a discussion of the quantitative time statistics regarding the creation of the bug reports from the known bug videos using both FUSION and the GCIT, then we examine the quantitative responses to the user experience questions and summarize the qualitative user preference responses.  

\begin{figure*}[tb]
\begin{center}
\includegraphics[width=\linewidth]{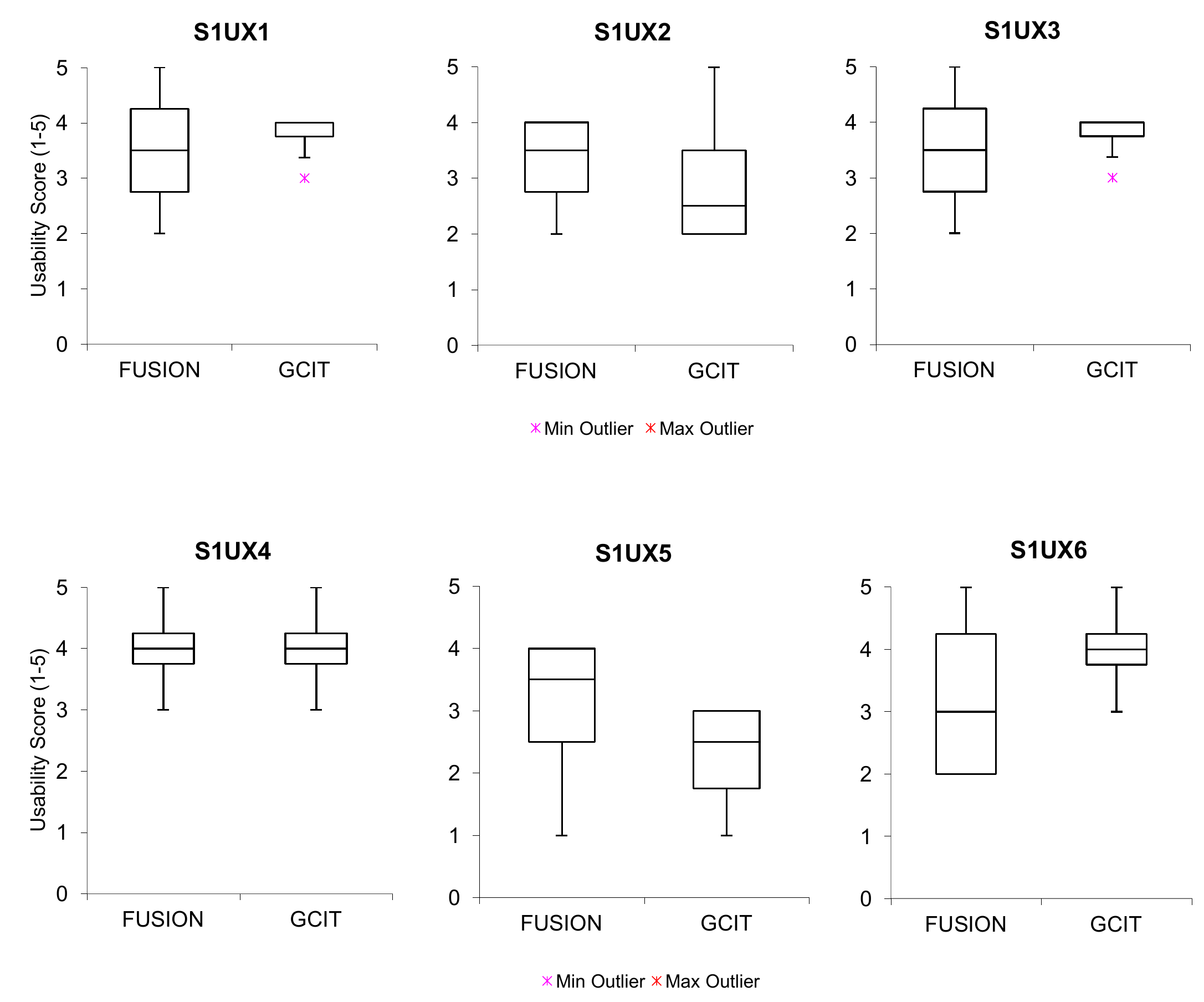}
\caption{\textbf{Study 1 User Experience Question Results:} Answers to the UX-related questions for Study 1 (Bug Report Creation)}
\label{fig:s1ux}
\end{center}
\end{figure*}
	
\begin{table}[p]
\centering
\small
\caption{\textbf{Creation Time Statistics for FUSION Bugs:} All of the times reported in this table are in the format (m:ss); an (*) next to the time indicates that FUSION was able to capture all of the steps for reproduction with the autocompletion engine and a replayable can be generated.}
\label{tab:fusion_creation_times}
\begin{tabular}{ | l | l | p{60pt} | p{60pt} | p{60pt} | p{60pt} | }
\hline
	Bug ID & App & Participant \#1 (Experienced) & Participant \#2 (Experienced) & Participant \#3 (Inexperienced) & Participant \#4 (Inexperienced) \\ \hline
	1 & A Time Tracker & 7:48 & 11:30 & 24:30 & 2:01 \\ \hline
	2 & Aarddict & 4:12 & 4:10 & 3:30 & 4:51 \\ \hline
	3 & ACV & 2:27 & 5:30 & 8:18 & 05:14 \\ \hline
	4 & Car Report & 12:21 & \textbf{4:50*} & 15:45 & \textbf{8:00*} \\ \hline
	5 & Document Viewer & \textbf{4:03*} & 5:10 & \textbf{16:32*} & \textbf{6:38*} \\ \hline
	6 & Droid Weight & \textbf{3:10*} & \textbf{2:10*} & \textbf{7:43*} & 6:09 \\ \hline
	7 & Eshotroid & 7:30 & 6:30 & 10:29 & 6:21 \\ \hline
	8 & GnuCash & 9:45 & \textbf{7:10*} & 18:45 & 08:23 \\ \hline
	9 & GnuCash & 9:23 & 7:30 & 20:03 & 9:27 \\ \hline
	10 & Mileage &  \textbf{2:22*} & 5:10 & 7:07 & \textbf{3:04*}\\ \hline
	11 & NetMBuddy & 2:02 & 3:15 & 4:00 & 1:27\\ \hline
	12 & Notepad & 3:53 & 3:20 & 4:45 & 3:14\\ \hline
	13 & OI Notepad &  5:15 & 9:20 & 13:30 & 6:17\\ \hline
	14 & Olam & 1:23 & 2:20 & 2:30 & 1:40 \\ \hline
	15 & QuickDic & 2:58 & 2:10 & 2:40 & 2:01\\ \hline
	 & \textbf{Average} & \textbf{5:14} & \textbf{5:20} & \textbf{10:40} & \textbf{4:59}\\ \hline
\end{tabular}
\end{table}

\begin{table}[p]
\centering
\small
\caption{\textbf{Creation Time Statistics for GCIT Bugs: } All of the times reported in this table are in the format (m:ss)}
\label{tab:gcit_creation_times}
\begin{tabular}{ | l | l | p{60pt} | p{60pt} | p{60pt} | p{60pt} | }
\hline
	Bug ID & App & Participant \#1 (Experienced) & Participant \#2 (Experienced) & Participant \#3 (Inexperienced) & Participant \#4 (Inexperienced) \\ \hline
	1 & A Time Tracker & 4:16 & 7:30 & 1:51 & 1:56 \\ \hline
	2 & Aarddict & 3:33 & 8:25 & 2:13 & 2:22\\ \hline
	3 & ACV &  2:37 & 11:10 & 0:51 & 1:42\\ \hline
	4 & Car Report & 2:52 & 12:23 & 0:40 & 2:39 \\ \hline
	5 & Document Viewer & 3:15 & 9:31 & 0:45 & 1:46 \\ \hline
	6 & Droid Weight & 2:33 & 7:13 & 1:03 & 1:45 \\ \hline
	7 & Eshotroid & 2:08 & 5:27 & 1:47 & 1:03 \\ \hline
	8 & GnuCash & 2:40 & 6:48 & 1:15 & 2:30 \\ \hline
	9 & GnuCash & 6:20 & 5:12 & 1:40 & 2:22 \\ \hline
	10 & Mileage & 3:53 & 5:25 & 1:00 & 1:16 \\ \hline
	11 & NetMBuddy & 3:52 & 3:13 & 1:20 & 1:48 \\ \hline
	12 & Notepad & 2:02 & 4:32 & 1:01 & 1:23 \\ \hline
	13 & OI Notepad & 3:16 & 6:25 & 0:58 & 1:12 \\ \hline
	14 & Olam & 4:26 & 3:13 & 1:16 & 1:49 \\ \hline
	15 & QuickDic & 1:37 & 03:17 & 0:55 & 0:59 \\ \hline
	 & \textbf{Average} & \textbf{3:17} & \textbf{6:39} & \textbf{1:14} & \textbf{1:46} \\ \hline
\end{tabular}
\end{table}

\subsection{Study 1 Bug Creation Time Results}
\label{study1_time}

Complete results for the bug report creation time statistics for Study 1 can be found in Tables \ref{tab:fusion_creation_times} and \ref{tab:gcit_creation_times}.  This data was collected during Study 1 in order to help quantify \textbf{RQ$_2$}, which aims to answer if FUSION is easier to use than traditional bug tracking systems for reporting bugs.  The length of time that a reporter spent filling out a bug report is an important indicator of the ease of use of the system.  The results of collecting this data show a clear trend, it took both experienced and inexperienced participants a longer amount of time to report bugs using the FUSION interface as compared to the GCIT, with the the total average bug creation time for FUSION being 6:33, compared to the total average bug creation time for the GCIT being 3:14.  However, there are also definite trends unique to each type of participant (e.g. experienced or inexperienced).  In particular, it is clear that there is a much smaller disparity in the time taken to complete bug reports for either system for the participants with prior programming experience, in fact it took one experienced participant longer to fill out bug reports for the GCIT than for FUSION.  This result is not unexpected, as a reporter with prior programming experience would be able to more easily navigate FUSION's UI and would also be more likely to more throughly enter information into a traditional bug tracking system such as the GCIT.  While the experienced participants showed little disparity in the creation times between the two reporting methods, inexperienced participants showed a very disparity, with the GCIT taking as much as 9 minutes faster on average.  This signifies that the Inexperienced users typically had more difficulty using the FUSION reporting system and entered only very brief natural language descriptions into the GCIT.  These results are not surprising, as experienced reporters understand the importance of providing detailed information in bug reports and thus are more likely to create detailed natural language bug reports using \textit{both} GCIT and FUSION.  On the other hand, the results show inexperienced reporters are more likely to create superficial reports using GCIT.  While it did take inexperienced reporters a longer amount of time to create FUSION reports, the creation times were still reasonable and doesn't necessarily reflect poorly on the system.  In fact, these results suggest that FUSION forced even inexperienced reporters to create more detailed, reproducible bug reports, and this is confirmed in the reproduction results. Furthermore, it is clear from responses to the user preferences questions that several users appreciated the structured nature, but would have preferred an improved web UI.  For instance, one participant stated: "In my opinion, the GUI component selector should not show the options as a list but in a easier way (for example a window where [you] pick the components)." These results contribute to the answer for \textbf{RQ$_2$} as follows: 
\newline
\newline
\noindent\fbox{%
    \parbox{\textwidth}{%
        \textbf{RQ$_2$:  The quantitive bug report creation time results suggest that FUSION is about as easy for developers to use as a traditional bug tracking system, however, it is more difficult for inexperienced users to use than traditional bug tracking systems.}
    }%
}

\subsection{Study 1 UX \& UP Results}
\label{study1_uxup}

In regard to the general usefulness of FUSION as tool for reporting bugs, there are two clear trends that emerge from the user responses: 1) \textit{Reporters generally feel that the opportunity to enter extra information in the form of detailed reproduction steps helps them more effectively report bugs}; 2) \textit{Experienced reporters tended to appreciate the value and added effort of adding extra information compared to inexperienced reporters.}  There are several statements made by participants that confirm these claims.  To highlight our first point, the responses we received were highly encouraging for S1UP6, for instance, on response read \textit{``With some small adjustments in the page (as I said, for example the GUI component selector) this system could really overtake all the existing bug-tracker systems in terms of usability and precision/detail of the generated bug report."} To highlight our second point made  One response to question S1UP1 from an experienced user was the following: ``The GUI component form and the action/event form have been very useful to effectively report the steps."; however a response to the same question by an inexperienced reporter was,``I liked the parts where you just type in the information".  However, these responses are not surprising, as the participants with programming experience understand the need for entering detailed information, but the inexperienced participants do not. One encouraging result during Study 1 is that FUSION was able to auto suggest all of the reproduction steps without gaps (i.e., auto-completion did not miss any steps) in 11 of 60 bug reports generated, as indicated in Table \ref{tab:fusion_creation_times}.  This means that, using the information for the steps contained with FUSION database, a replayable script can be generated, whereas this would not be possible for GCIT or any other bug tracking system. 
	The user experience statistics from \textit{Study 1} are listed in Figure \ref{fig:s1ux}.  In this table questions 1,3,4, and 6 expect an answer of Strongly Agree (5) in order to correlate to a favorable usability score.  Experienced developers reported scores of 4.5 for each of these questions, indicating that developers give FUSION a high usability score.  Inexperienced users gave the same questions scores of 2.5, 2.5, 3.5, and 2 respectively, indicating a low usability score.  Thus, the results show two major trends: 1) \textit{Experienced users tended to prefer FUSION compared to the GCIT} and 2) \textit{Inexperienced users tended to prefer the GCIT compared to FUSION}.  It should be noted, that since there are only two experienced and inexperienced respondents for each system, the results in this section are not generalizable.  These results are not surprising, as \textit{non-expert} users seemed to prefer the simplicity of the Natural Language text entry of the GCIT to the structured format of FUSION.  In summary we can answer \textbf{RQ$_1$} as follows: 
\newline
\newline
\noindent\fbox{%
    \parbox{\textwidth}{%
        \textbf{RQ$_1$:  While reporter's generally felt that the opportunity to enter extra information in a bug report using FUSION increased the quality of their reports, inexperienced users would have preferred a simpler web UI.}
    }%
}

\section{Study 2 (Bug Report Reproducibility) Results}
\label{results2}

In this section we present the qualitative and quantitative results for \textit{Empirical Study 2}. We begin with a discussion of the  the quantitative responses to the user experience questions and summarize the qualitative user preference responses. Then we examine quantitative time and reproducibility statistics regarding the reproduction of bugs from FUSION, GCIT, and Original bug reports.

\subsection{Study 2 UX and UP Results}
\label{study2_uxup}

\begin{figure*}[h!]
\begin{center}
\includegraphics[width=0.9\linewidth]{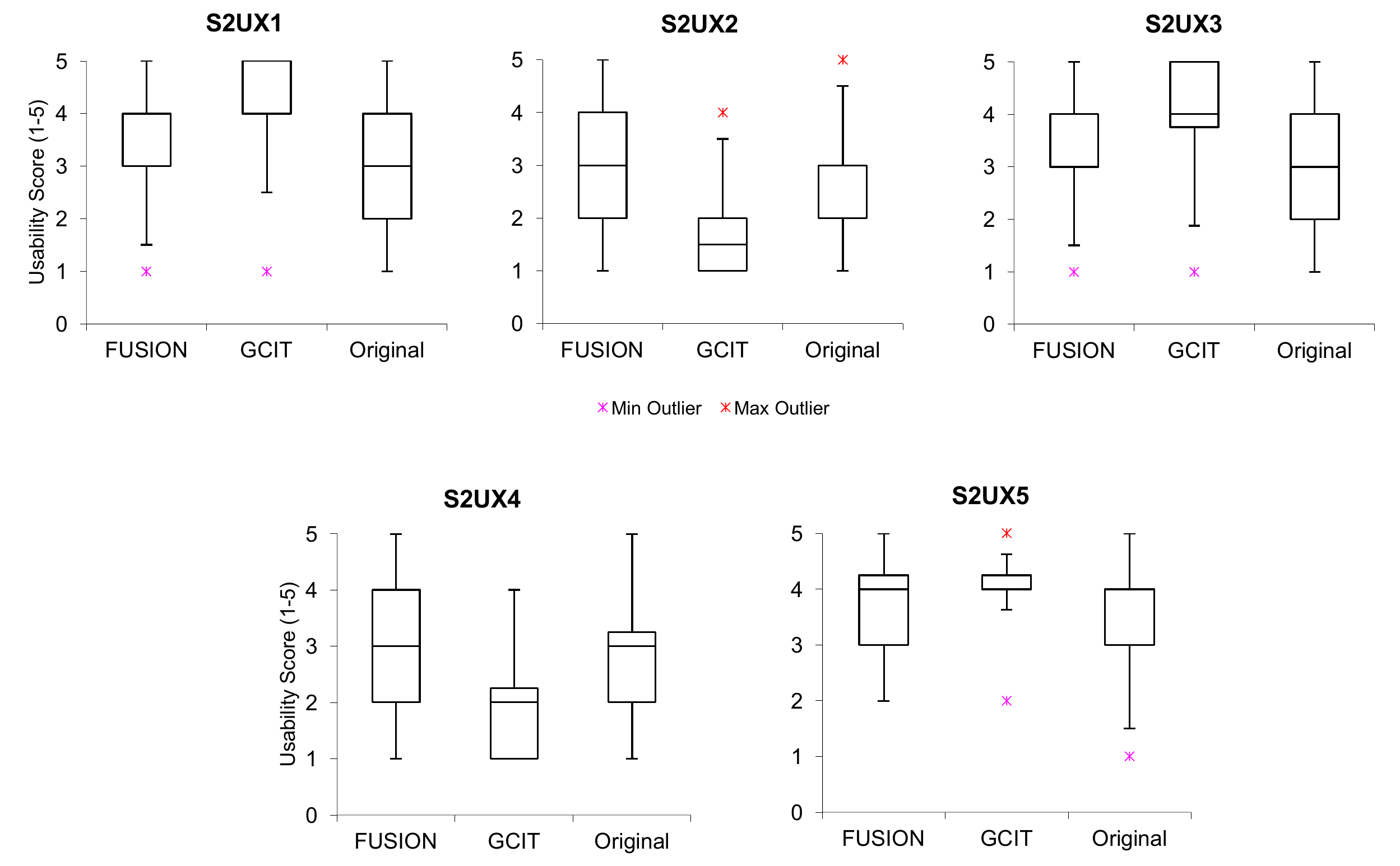}
\caption{\textbf{Study 2 User Experience Question Results:} Answers to the UX-related questions for Study 2 (Bug Report Reproduction)}
\label{fig:s2ux}
\end{center}
\end{figure*}

  The usability scores in Figure \ref{fig:s2ux} show that most users agree that they would like to use FUSION's bug reports frequently, however, several users also found the bug reports to be unnecessarily complex, and some users found the bug reports difficult to read/comprehend.  Most users agreed that they thought FUSION bug reports were useful for helping to reproduce the bugs.  GCIT had the best usability scores out of the three systems, whereas the Original bug reports had the lowest usability scores.  According to user preference feedback which asked what information participants found most useful in bug reports we received encouraging feedback; for instance: ``The detail steps to find where to find the next steps was really useful and speeded up things."; ``The images of icons help a lot, especially when you have a hard time locating the icons on your screen.".  However, users also expressed issues with the FUSION report layout: ``Sometimes the steps were too overly specific/detailed."; ``The information, while thorough, was not always clear"; ``If there are steps missing, it is confusing because it is otherwise so detailed".  Based on these responses we can answer \textbf{RQ$_2$} as follows: 
  \newline
\newline
\noindent\fbox{%
    \parbox{\textwidth}{%
        \textbf{RQ$_2$: According to usability scores, participants generally preferred FUSION over the original bug reports, but generally preferred GCIT to FUSION by a small margin.  The biggest reporter complaint regarding FUSION was the organization of information in the report. }
    }%
}

\subsection{Study 2 Bug Reproduction Results}
\label{study2_repo}

\begin{table}[t]
\centering
\small
\caption{\textbf{Average Bug Report Reproduction Time:} Average reproduction time results for each type of bug report evaluated.}
\vspace{+0.3cm}
\label{avg-time}
\begin{tabular}{ | l | c |}
\hline
\textbf{Bug Report Type} &  \textbf{ Avg Time to Reproduce} \\ \hline
FUSION (E) & 3:15 \\ \hline 
FUSION(I) & 2:35 \\ \hline 
Google Code (E) & 1:46 \\ \hline 
Google Code (I) & 1:46 \\ \hline 
Originial & 1:59 \\ \hline 
FUSION Total & 2:55\\ \hline
Google Code Total & 1:46 \\ \hline
\end{tabular}
\end{table}

\begin{figure}[tb]
\begin{center}
\includegraphics[width=0.9\linewidth]{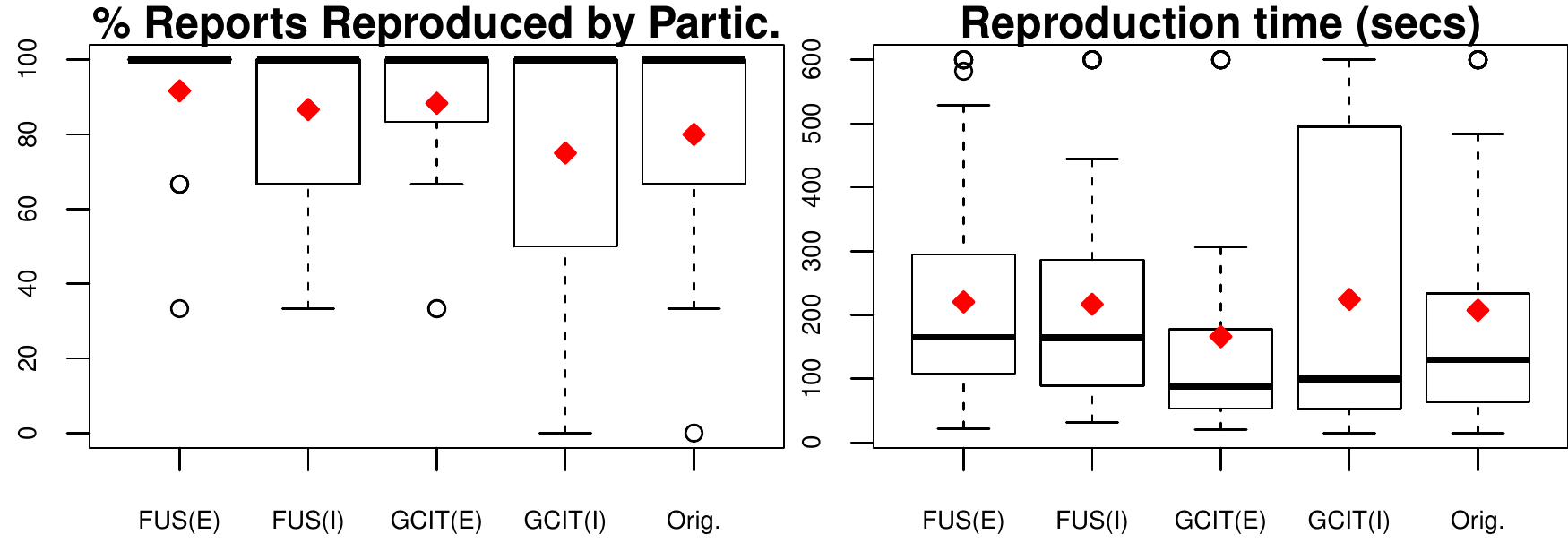}
\caption{\textbf{Study 2 Bug Report Reproduction Results:} Results for the number of bug reports reproduced and the average time taken to reproduce each bug}
\label{fig:repo_results}
\end{center}
\end{figure}

The Boxplots in Figure \ref{fig:repo_results} summarize the reproduction results for \textit{Study 2}, and more detailed statistics about the number of bugs that could not be reproduced for each system, as well as average reproduction times for each type of report can be found in Tables \ref{repo} and \ref{avg-time} respectively.  In the case of reproduction time, because some of the reports were not reproduced during a 10 minutes time slot, we set to 600 seconds the reproduction time for visualization and analysis purposes. Detailed results regarding the reproduction of bug reports can be found in Table \ref{tab:p1-reproduction}.

As mentioned earlier there are five types of bug reports that this study evaluates: FUSION reports written by experienced (i.e., FUSE(E)) and non-experienced participants (i.e., FUS(I)), reports written in GCIT by experienced (i.e., GCIT(E)) and non-experienced participants (i.e., GCIT(I)), and original reports (i.e., Orig). As mentioned before, for the analysis we will assume that the reproduction time of the non-reproducible bug reports is the maximum time the participants had to declare a report as reproducible or not. This decision is to have fair comparisons and avoid bias towards the options with a low rate of non-reproducible reports (e.g., FUS(E)). Consequently, The average time to reproduce for the two flavors of FUSION were 220.5 seconds and 216.8 seconds seconds respectively for FUS(E) and FUS(I).  Surprisingly, the FUS(I) reports had a smaller average time to reproduce than the FUS(E) reports.  Both types of GCIT reports (E) \& (I) had an average time to reproduce of 166.07 seconds and 224.45 seconds. While this result shows that participants took longer to reproduce FUSION reports, this is to be expected as they had to read and process the extra information regarding the reproduction steps. However, reproduction time of inexperienced reporters with  FUSION is lower than GCIT.  There is a clear trade-off of reproduction time versus accuracy.  When examining the effectiveness of the reports for bugs that were seemingly more complex to reproduce (e.g., they took more time overall to reproduce), we see there is no strong correlation between the relative effectiveness of FUSION or GCIT. However, we do see that the more complex bugs generally have more instances where they are not reproducible, which is to be expected. Based on these results we can answer {RQ$_3$} as follows: 
\newline
\newline
\noindent\fbox{%
    \parbox{\textwidth}{%
        \textbf{RQ$_3$:  Bug reports generated with FUSION do not allow for faster reproduction of bugs compared bug reports generated using traditional bug tracking systems such as the GCIT}
    }%
}
\newline

Figure \ref{fig:repo_results} details reproducibility results for bug reports written in FUSION. In terms of reproducibility, overall the reports generated using FUSION were more reproducible than the reports generated using GCIT with only 13 of the 120 bug reports from FUSION being non-reproducible compared to 23 of the 120 reports from the GCIT being non-reproducible. The bug report type with lowest number of non-reproducible cases is FUS(E), where as the bug report type with the highest number of non-reproducible cases is GCIT(I)  One encouraging result is that when inexperienced participants created bug reports in \textit{Study 1}, participants in \textit{Study 2} seemed to have a much easier time reproducing the reports from FUSION (I) which only had eight non-reproducible cases, compared to GCIT(I) which had twice as many, 15, non-reproducible cases. This means that for reporters that may be classified as inexperienced  FUSION could greatly improve the bug report quality.  Both of the individual FUSION bug report types (I) and (E) had a lower number of non-reproducible cases than the Original bug reports as well.  However, a direct comparison cannot be made here, as each original bug report was tested four times, compared to two times for the FUSION and GCIT based bug reports.  Therefore, based on these results we can answer \textbf{RQ$_4$} as follows: 
\newline
\newline
\noindent\fbox{%
    \parbox{\textwidth}{%
        \textbf{RQ$_4$:  Developers using FUSION are able to reproduce more bugs compared to traditional bug tracking systems such as the GCIT.}
    }%
}
\newline

\begin{table}[tb]
\centering
\small
\caption{\textbf{Non Reproducible Bug Reports:} Number of bugs that could not be reproduced per Bug Report Type. }
\label{repo}
\begin{tabular}{ | l | c |}
\hline
\textbf{Bug Report Type} & \textbf{\# of Bugs that could not be reproduced} \\ \hline
FUSION (E) & 5\\ \hline
FUSION(I) & 8\\ \hline
Google Code (E) & 8\\ \hline 
Google Code (I) & 15\\ \hline 
Originial & 11\\ \hline 
FUSION Total & 13\\ \hline 
Google Code Total & 23 \\ \hline
\end{tabular}
\end{table}

\begin{sidewaystable*}[p]
\centering
\small
\caption{\textbf{Bug Report Quality Statistics:} GCIT = Google Code Issue Tracker, NR = \# Instances not reproducible, Time = Average time to reproduce, E = bug report created by experienced participant, I = bug report created by inexperienced participant}
\label{tab:p1-reproduction}
\begin{tabular}{ | p{54pt} | p{31pt} | p{33pt} | p{33pt} | p{31pt} | p{31pt} | p{31pt} | p{31pt} | p{31pt} | p{31pt} | p{31pt} | p{31pt} | }
\hline
	App&Total Average Time&FUSION (E) Time&FUSION (E) NR&FUSION (I) Time&FUSION (I) NR&GCIT (E) Time&GCIT (E) NR&GCIT (I) Time&GCIT (I) NR&Original Time& Original NR \\ \hline
	\textbf{Time Tracker} & 2:29 & 2:42 & 0 & 2:46 & 1 & 1:33 & 3 & 3:02 & 1 & 2:00 & 1 \\ \hline
	\textbf{Aarddict} & 2:53 & 4:25 & 1 & 4:34 & 1 & 1:31 & 1 & 1:13 & 1 & 2:43 & 2 \\ \hline
	\textbf{ACV} & 2:09 & 2:45 & 0 & 2:24 & 0 & 1:55 & 0 & 2:10 & 2 & 1:29 & 2 \\ \hline
	\textbf{Car Report} & 2:38 & 5:14 & 0 & 2:00 & 0 & 1:21 & 0 & N/A & 4 & 1:57 & 0 \\ \hline
	\textbf{Document Viewer} & 2:08 & 2:00 & 1 & 1:37 & 0 & 2:06 & 0 & 3:13 & 2 & 1:46 & 0 \\ \hline
	\textbf{DroidWeight} & 1:47 & 3:47 & 0 & 1:29 & 0 & 0:58 & 0 & 1:38 & 0 & 1:02 & 1 \\ \hline
	\textbf{Eshotroid} & 2:20 & 3:06 & 0 & 1:38 & 0 & 2:35 & 1 & 2:07 & 1 & 2:13 & 0 \\ \hline
	\textbf{GnuCash 1} & 3:57 & 5:17 & 1 & N/A & 3 & 3:58 & 1 & 1:33 & 2 & 5:01 & 1 \\ \hline
	\textbf{GnuCash 2} & 2:25 & 3:48 & 1 & 3:21 & 2 & 1:33 & 1 & 1:37 & 1 & 1:45 & 2 \\ \hline
	\textbf{Mileage} & 1:54 & 2:25 & 0 & 2:17 & 0 & 2:00 & 0 & 1:12 & 0 & 1:37 & 0 \\ \hline
	\textbf{NetMBuddy} & 1:35 & 2:19 & 0 & 1:34 & 0 & 0:51 & 0 & 1:02 & 0 & 2:10 & 0 \\ \hline
	\textbf{Notepad} & 1:38 & 1:40 & 0 & 3:35 & 1 & 1:05 & 0 & 0:51 & 0 & 1:00 & 0 \\ \hline
	\textbf{OI Notepad} & 4:14 & 6:11 & 0 & 5:22 & 0 & 3:25 & 0 & 3:47 & 0 & 2:26 & 1 \\ \hline
	\textbf{Olam} & 1:00 & 0:49 & 0 & 0:50 & 0 & 0:54 & 0 & 1:16 & 0 & 1:09 & 0 \\ \hline
	\textbf{QuickDic} & 2:05 & 3:34 & 0 & 2:36 & 0 & 1:36 & 0 & 0:56 & 0 & 1:41 & 0 \\ \hline \hline
	\textbf{Total/Avg}: &&3:15&5&2:35&8&1:46&8&1:46&15&1:59&11 \\ \hline
\end{tabular}
\end{sidewaystable*}
\chapter{Limitations and Threats to Validity}
\label{limitations}

\section{Limitations of the FUSION approach}

Currently, the DFS implementation in FUSION only supports the \textit{click/tap} action.  Another option to gather runtime program information would be to record app scenarios and replay them while collecting program data or using language modeling based approaches for scenario generation \cite{Linares:MSR15} .  However, we forwent these approaches in favor of the fully automatic DFS application exploration.  Part of our immediate plan for future work includes adding support for more gestures to our DFS engine.  FUSION is currently not capable of capturing certain contextual app  information such as a change in device orientation or network state.  However, this can be mitigated by the fact that reporters can enter such contextual information in the free-form text field associated with each step.  FUSION is also limited in the types of bugs that it can report.  For instance, certain performance or energy bugs would not be as useful reported through FUSION, as the steps to reduce for these non-functional types of bugs may not be as well defined as bugs that can be triggered by manipulating GUI components.  FUSION is also limited in the types of bugs that it can report, currently supporting functional bugs that can be uncovered using only GUI-Gestures such as tap, long-touch, swipe and type.  It is important to note that even though the systematic section engine is not able to perform and capture gestures other than tap, these gestures can still be reported using FUSION. Also, FUSION does not perform any analysis on stack traces in order to reverse-engineer reproduction steps, it is purely a mechanism to aid users in reporting functional, and GUI-related bugs.  

\section{Threats to the Validity of the Empirical Study}
\label{threats}
	
	Threats to internal validity concern issues with the validity of causal relationships inferred.  In the context of our studies, threats come from potentially confounding effects of participants. Since we base our conclusions upon data collected from participants, we see two major threats to validity.  First, we assumed that undergraduate students without a CS background, but those who had experience using Android devices are representative of \textit{non-expert} testers.  We believe this is a reasonable assumption given the context as most \textit{non-expert} testers will only have a ``working" knowledge of the app and platform.  We also assumed graduate students with Android experience were reasonable substitutes for developers.  Again, we believe this is a reasonable assumption given that all four of the ``experienced" participants in \textit{Study 1} indicated they had extensive programming backgrounds and reasonable Android programming experience (above 4 on the scale where 10 represents ``Very experienced").  Likewise, the participants in \textit{Study 2} indicated that they all had extensive programming backgrounds, and 13 of the 20 participants had reasonable Android programming experience.

Threats to external validity concern the generalizability of the results.  The first threat to the generalizability of the results relates to the concern of the bug reports and Android apps used in our study.  We evaluated FUSION on only 15 bug reports from 14 different applications from the F-droid \cite{fdroid} marketplace.  In order to increase the generalizability of the results we aimed at selecting bug reports of varying type and complexity from apps representing different categories and functions.  During our study we also utilized only one device type, a Nexus 7 tablet.  However, this was for the purpose of standardizing the results across all the participants.  There is nothing limiting us from using FUSION on many different Android devices from varied manufacturers.  We concede that FUSION may not necessarily be suited for reporting all types of bugs, (e.g., nuanced performance bugs), however, we conjecture that any type of bug that can be reported with a traditional issue tracking system can be reported with FUSION.
\chapter{Conclusion}
\label{conclusion}

Prior research has shown that the high-abstraction level of natural language descriptions in current bug tracking systems makes it difficult for reporters to provide actionable information to developers.  This illustrates the \textit{lexical gap} between reporters of bugs and developers.  To help overcome this gap, we introduced FUSION, a novel bug reporting approach that takes advantage of program analysis techniques and the event-driven nature of Android applications in order to help auto-complete the reproduction steps for bugs.  We evaluated FUSION on 15 real-world Android application bugs in a user study involving 28 participants and show that reports generated by FUSION are more reliable for producing bugs than reports from the issue tracking system integrated into Google-Code.  We hope our work on FUSION encourages a new direction of research regarding improving reporting systems.
	In the future, we aim to improve our DFS engine through supporting gestures, to explore adding more specific program information in reports for quicker/automatic fault localization, and to use FUSION as a tool for reporting feature requests to aid feature location \cite{Posyvanyk:ICPC07,Poshyvanyk:TSEM13,Poshyvanyk:ICPC06,Poshyvanyk:TSE07,Liu:ASE07,Revelle:ICPC10,Dit:JSEP13,Dit:ESE13} and impact analysis tasks \cite{Dit:ICSE14,Poshyvanyk:ESE09,Kagdi:WCRE10,Gethers:ICSE12,Kagdi:ESE13}.			

\appendix
\chapter{Instructions for User Study Participants}
\label{appen:user_instrucs}

\section{Study 1 Instructions}
\label{s1_instructions}

Thank you for agreeing to participate in my user study, I sincerely appreciate your assistance. Before starting the study, please be sure to read and sign the consent form that I will give you at the beginning of the study. The study you are participating in today has two tasks. Your first task will be to recreate a bug demonstrated in an online video on a physical Nexus 7 tablet device that will be provided to you for the duration of the study. The second task will be to fill out a bug report, and time yourself while doing this, for each of the bugs that you have recreated on a tablet, in a specified Bug Tracking System, System A. For those of you who may not be familiar with the term ?bug? or ?bug tracking system? please see the following definitions:
A \textbf{software bug}\cite{wiki-bug} is an error, flaw, failure, or fault in a computer program or system that causes it to produce an incorrect or unexpected result, or to behave in unintended ways.
\textbf{bug tracking system - (BTS):}\cite{wiki-bts} A system for receiving and filing bugs reported against a software project, and tracking those bugs until they are fixed. Most major software projects have their own BTS, the source code of which is often available for use by other projects.
You will be asked to record your participation number, the unique id of each bug report you fill out, and the time it takes you complete each bug report, and some exit interview questions after completing your task. The link for the survey can be found here: \url{https://www.surveymonkey.com/s/System_A_Feedback} You will time yourself starting from the point you first start filling out each bug report, until the time you hit the submit button when you have finished filling out each bug report. The steps you should take for each bug report are as follows:
\begin{enumerate}
\item Watch the video for the bug in question. (Each video has a description and lists the App to which it corresponds)
\item Attempt to reproduce the bug on the tablet device loaned to you (You will find all of the apps for the user study in a folder on the Home screen titled ?User Study?. You should verify with Kevin that you have successfully recreated each bug)
\item Open the link to the bug tracking system here: \url{http://23.92.18.210:8080/FusionWeb} and select the app for which you are filling out the bug report.
\item Open up the survey (if you haven?t already) and copy and paste the unique id for the bug report that you are filling out into the survey.
\item Start your timer and fill out the bug report to the best of your ability.
\item When you hit the submit button, signifying that you have finished entering the bug report, stop your timer and record the time next to the corresponding bug report id in the survey.
\item Ensure the information you have entered into the survey is correct, and start the process over with the next bug in the list.
\item If have a question at any point during the study, please ask and I will do my best to answer.
\end{enumerate}

\section{Study 2 Instructions}
\label{s2_instructions}

Thank you for agreeing to participate in this user study, I sincerely appreciate your assistance. Before starting, please be sure to read and sign the consent form that you will be given before the the proctor gives an overview of the tasks. The study you are participating in today has two tasks. Your first task will be to attempt to reproduce fifteen different bugs from three different types of bug reports on a Google Nexus 7 tablet that will be provided to you for the duration of the study. You can find the links to the fifteen bugs assigned to you below. For each bug, you should time yourself from the point you open the corresponding bug report until you perform the last step that manifests the bug. When you think you have correctly reproduced the bug, please call over Kevin and he will either verify that you have successfully reproduced the bug, or tell you to keep trying. If Kevin confirms that you successfully reproduced the bug, record your time in the first page of the survey (\url{https://www.surveymonkey.com/s/ Bug_Reproduction_Survey}) in the corresponding bug?s text box. There is a ten minute time limit for reproducing each bug, if you cannot reproduce a bug within ten minutes, stop, record the 10 minute time in the first page of the survey and move on to the next bug. Your second task will be to answer some questions regarding each type of bug report you encountered as well as some demographic questions in
an online survey.
To begin, please open the survey and enter your participation number (number only) and your
department affiliation as well as the current degree you are pursuing in the first screen of the survey. Then follow the steps below to complete your tasks:
Steps to Complete the Study:
\begin{enumerate}
\item Open the link to the bug that you wish to reproduce. (You will find the app name next to the link for each bug report, and all of the apps for the study are in a folder on the home screen)
\item Start your timer and carefully read the bug report.
\item Open the app on the Nexus 7 tablet, and attempt to reproduce the bug from the bug report.
\item Once you fell you have reproduced the bug, stop your timer, call Kevin over, and he will verify wether or not you have correctly reproduced the bug.
\item If you did correctly reproduce the bug, record the time it took you on the first page of the survey, and move on to the next bug in the list. If you did not successfully reproduce the bug, keep trying either until Kevin verifies that you succeeded, or until the ten minute time limit, then enter your time result into the survey and move on to the next bug in the list.
\item After you have attempted to reproduce all bugs, please answer the survey questions as thoroughly and honestly as possible, verify your results have been received with Kevin before leaving the study.
\end{enumerate}

** If at any point in the study you have a questions please ask and the proctor will do their best to answer, however, there will be some questions that we are not able to answer due to constraints of the investigation. **

\makeThesisBib{ms.bib}

\begin{thebibliography}{10}

\bibitem{uiautomator}
Android uiautomator
  \url{http://developer.android.com/tools/help/uiautomator/index.html}.

\bibitem{apktool}
apktool \url{https://code.google.com/p/android-apktool/}.

\bibitem{wiki-bts}
Atimetrackerbug \url{https://en.wikipedia.org/wiki/Bug_tracking_system/}.

\bibitem{wiki-bug}
Atimetrackerbug \url{https://en.wikipedia.org/wiki/Software_bug}.

\bibitem{bugzilla}
Bugzilla issue tracker \url{https://bugzilla.mozilla.org}.

\bibitem{dex2jar}
dex2jar \url{https://code.google.com/p/dex2jar/}.

\bibitem{fdroid}
F-droid. \url{https://f-droid.org/}.

\bibitem{github-it}
Github issue tracker \url{https://github.com/features}.

\bibitem{google-code}
Google code issue tracker
  \url{https://code.google.com/p/support/wiki/IssueTracker}.

\bibitem{jd-cmd}
jd-cmd decompiler \url{https://github.com/kwart/jd-cmd}.

\bibitem{jira}
Jira bug reporting system \url{https://www.atlassian.com/software/jira}.

\bibitem{mantis}
Mantis bug reporting system \url{https://www.mantisbt.org}.

\bibitem{app-abandonment}
Mobile apps: What consumers really need and want
  \url{https://info.dynatrace.com/rs/compuware/images/Mobile_App_Survey_Report.pdf}.

\bibitem{srcml}
srcml \url{http://www.srcml.org}.

\bibitem{usersnap}
Usersnap bug reorting tool
  \url{https://usersnap.com/features/feedback-widget-for-screenshot-bug-reporting}.

\bibitem{55Amalfitano:ASE2012}
{\sc Domenico Amalfitano, Anna~Rita Fasolino, Porfirio Tramontana, Salvatore
  De~Carmine, and Atif~M. Memon}.
\newblock Using gui ripping for automated testing of android applications.
\newblock In {\em Proceedings of the 27th IEEE/ACM International Conference on
  Automated Software Engineering}, ASE 2012, pages 258--261, New York, NY, USA,
  2012. ACM.

\bibitem{Amalfitano:ASE2012}
{\sc Domenico Amalfitano, Anna~Rita Fasolino, Porfirio Tramontana, Salvatore
  De~Carmine, and Atif~M. Memon}.
\newblock Using gui ripping for automated testing of android applications.
\newblock In {\em Proceedings of the 27th IEEE/ACM International Conference on
  Automated Software Engineering}, ASE 2012, pages 258--261, New York, NY, USA,
  2012. ACM.

\bibitem{34Aranda:ICSE09}
{\sc J.~Aranda and G.~Venolia}.
\newblock The secret life of bugs: Going past the errors and omissions in
  software repositories.
\newblock In {\em Software Engineering, 2009. ICSE 2009. IEEE 31st
  International Conference on}, pages 298--308, May 2009.

\bibitem{41Artzi:ECOOP2008}
{\sc Shay Artzi, Sunghun Kim, and MichaelD. Ernst}.
\newblock Recrash: Making software failures reproducible by preserving object
  states.
\newblock In {\em ECOOP 2008 -- Object-Oriented Programming}, Jan Vitek,
  editor, volume 5142 of {\em Lecture Notes in Computer Science}, pages
  542--565. Springer Berlin Heidelberg, 2008.

\bibitem{49Ayewah:IS2008}
{\sc N.~Ayewah, D.~Hovemeyer, J.D. Morgenthaler, J.~Penix, and William Pugh}.
\newblock Using static analysis to find bugs.
\newblock {\em Software, IEEE}, 25(5):22--29, Sept 2008.

\bibitem{Azim:OOPSLA2013}
{\sc Tanzirul Azim and Iulian Neamtiu}.
\newblock Targeted and depth-first exploration for systematic testing of
  android apps.
\newblock In {\em Proceedings of the 2013 ACM SIGPLAN International Conference
  on Object Oriented Programming Systems Languages \&\#38; Applications},
  OOPSLA '13, pages 641--660, New York, NY, USA, 2013. ACM.

\bibitem{38Baudry:ICSE2006}
{\sc Benoit Baudry, Franck Fleurey, and Yves Le~Traon}.
\newblock Improving test suites for efficient fault localization.
\newblock In {\em Proceedings of the 28th International Conference on Software
  Engineering}, ICSE '06, pages 82--91, New York, NY, USA, 2006. ACM.

\bibitem{27Bell:ICSE13}
{\sc Jonathan Bell, Nikhil Sarda, and Gail Kaiser}.
\newblock Chronicler: Lightweight recording to reproduce field failures.
\newblock In {\em Proceedings of the 2013 International Conference on Software
  Engineering}, ICSE '13, pages 362--371, Piscataway, NJ, USA, 2013. IEEE
  Press.

\bibitem{3Bettenburg:FSE08}
{\sc Nicolas Bettenburg, Sascha Just, Adrian Schr\"{o}ter, Cathrin Weiss, Rahul
  Premraj, and Thomas Zimmermann}.
\newblock What makes a good bug report?
\newblock In {\em Proceedings of the 16th ACM SIGSOFT International Symposium
  on Foundations of Software Engineering}, SIGSOFT '08/FSE-16, pages 308--318,
  New York, NY, USA, 2008. ACM.

\bibitem{32Bettenburg:ICSM08}
{\sc Nicolas Bettenburg, R.~Premraj, T.~Zimmermann, and Sunghun Kim}.
\newblock Duplicate bug reports considered harmful... really?
\newblock In {\em Software Maintenance, 2008. ICSM 2008. IEEE International
  Conference on}, pages 337--345, Sept 2008.

\bibitem{11Bettenburg:MSR08}
{\sc Nicolas Bettenburg, Rahul Premraj, Thomas Zimmermann, and Sunghun Kim}.
\newblock Extracting structural information from bug reports.
\newblock In {\em Proceedings of the 2008 International Working Conference on
  Mining Software Repositories}, MSR '08, pages 27--30, New York, NY, USA,
  2008. ACM.

\bibitem{5Bhattacharya:CSMR13}
{\sc P.~Bhattacharya, L.~Ulanova, I.~Neamtiu, and S.C. Koduru}.
\newblock An empirical analysis of bug reports and bug fixing in open source
  android apps.
\newblock In {\em Software Maintenance and Reengineering (CSMR), 2013 17th
  European Conference on}, pages 133--143, March 2013.

\bibitem{15Breu:CSCW10}
{\sc Silvia Breu, Rahul Premraj, Jonathan Sillito, and Thomas Zimmermann}.
\newblock Information needs in bug reports: Improving cooperation between
  developers and users.
\newblock In {\em Proceedings of the 2010 ACM Conference on Computer Supported
  Cooperative Work}, CSCW '10, pages 301--310, New York, NY, USA, 2010. ACM.

\bibitem{Brooke:96}
{\sc J.~Brooke}.
\newblock {SUS}: A quick and dirty usability scale.
\newblock In {\em Usability evaluation in industry}, P.~W. Jordan,
  B.~Weerdmeester, A.~Thomas, and I.~L. Mclelland, editors. Taylor and Francis,
  London, 1996.

\bibitem{43Cao:ASE14}
{\sc Yu~Cao, Hongyu Zhang, and Sun Ding}.
\newblock Symcrash: Selective recording for reproducing crashes.
\newblock In {\em Proceedings of the 29th ACM/IEEE International Conference on
  Automated Software Engineering}, ASE '14, pages 791--802, New York, NY, USA,
  2014. ACM.

\bibitem{Chen:icse2014}
{\sc N.~Chen, J.~Lin, S.~Hoi, X.~Xiao, and B.~Zhang}.
\newblock {AR-Miner}: Mining informative reviews for developers from mobile app
  marketplace.
\newblock In {\em 36th International Conference on Software Engineering
  (ICSE'14)}, page To appear, 2014.

\bibitem{Choi:OOPSLA2013}
{\sc Wontae Choi, George Necula, and Koushik Sen}.
\newblock Guided gui testing of android apps with minimal restart and
  approximate learning.
\newblock In {\em Proceedings of the 2013 ACM SIGPLAN International Conference
  on Object Oriented Programming Systems Languages \&\#38; Applications},
  OOPSLA '13, pages 623--640, New York, NY, USA, 2013. ACM.

\bibitem{29Clause:ICSE07}
{\sc James Clause and Alessandro Orso}.
\newblock A technique for enabling and supporting debugging of field failures.
\newblock In {\em Proceedings of the 29th International Conference on Software
  Engineering}, ICSE '07, pages 261--270, Washington, DC, USA, 2007. IEEE
  Computer Society.

\bibitem{52Cleve:ICSE2005}
{\sc Holger Cleve and Andreas Zeller}.
\newblock Locating causes of program failures.
\newblock In {\em Proceedings of the 27th International Conference on Software
  Engineering}, ICSE '05, pages 342--351, New York, NY, USA, 2005. ACM.

\bibitem{Czarnecki:ICSM2012}
{\sc Krzysztof Czarnecki, Zeeshan Malik, and Rafael Lotufo}.
\newblock Modelling the \&\#8216;hurried\&\#8217; bug report reading process to
  summarize bug reports.
\newblock In {\em Proceedings of the 2012 IEEE International Conference on
  Software Maintenance (ICSM)}, ICSM '12, pages 430--439, Washington, DC, USA,
  2012. IEEE Computer Society.

\bibitem{54Dallmeier:LNCS2005}
{\sc Valentin Dallmeier, Christian Lindig, and Andreas Zeller}.
\newblock Lightweight defect localization for java.
\newblock In {\em ECOOP 2005 - Object-Oriented Programming}, AndrewP. Black,
  editor, volume 3586 of {\em Lecture Notes in Computer Science}, pages
  528--550. Springer Berlin Heidelberg, 2005.

\bibitem{31Davies:ESEM2014}
{\sc Steven Davies and Marc Roper}.
\newblock What's in a bug report?
\newblock In {\em Proceedings of the 8th ACM/IEEE International Symposium on
  Empirical Software Engineering and Measurement}, ESEM '14, pages 26:1--26:10,
  New York, NY, USA, 2014. ACM.

\bibitem{Dit:JSEP13}
{\sc Bogdan Dit, Meghan Revelle, Malcom Gethers, and Denys Poshyvanyk}.
\newblock Feature location in source code: a taxonomy and survey.
\newblock {\em Journal of Software: Evolution and Process}, 25(1):53--95, 2013.

\bibitem{Dit:ESE13}
{\sc Bogdan Dit, Meghan Revelle, and Denys Poshyvanyk}.
\newblock Integrating information retrieval, execution and link analysis
  algorithms to improve feature location in software.
\newblock {\em Empirical Softw. Engg.}, 18(2):277--309, April 2013.

\bibitem{Dit:ICSE14}
{\sc Bogdan Dit, Michael Wagner, Shasha Wen, Weilin Wang, Mario
  Linares-V\'{a}squez, Denys Poshyvanyk, and Huzefa Kagdi}.
\newblock Impactminer: A tool for change impact analysis.
\newblock In {\em Companion Proceedings of the 36th International Conference on
  Software Engineering}, ICSE Companion 2014, pages 540--543, New York, NY,
  USA, 2014. ACM.

\bibitem{4Joorabchi:MSR14}
{\sc Mona Erfani~Joorabchi, Mehdi Mirzaaghaei, and Ali Mesbah}.
\newblock Works for me! characterizing non-reproducible bug reports.
\newblock In {\em Proceedings of the 11th Working Conference on Mining Software
  Repositories}, MSR 2014, pages 62--71, New York, NY, USA, 2014. ACM.

\bibitem{24MobilityReport}
{\sc Ericsson}.
\newblock Ericsson mobility report novmeber 2014.
\newblock
  http://www.ericsson.com/res/docs/2014/ericsson-mobility-report-november-2014.pdf,
  November 2014.

\bibitem{Feigenspan:ICPC12}
{\sc J.~Feigenspan, C.~Kastner, J.~Liebig, S.~Apel, and S.~Hanenberg}.
\newblock Measuring programming experience.
\newblock In {\em Program Comprehension (ICPC), 2012 IEEE 20th International
  Conference on}, pages 73--82, June 2012.

\bibitem{Gethers:ICSE12}
{\sc Malcom Gethers, Bogdan Dit, Huzefa Kagdi, and Denys Poshyvanyk}.
\newblock Integrated impact analysis for managing software changes.
\newblock In {\em Proceedings of the 34th International Conference on Software
  Engineering}, ICSE '12, pages 430--440, Piscataway, NJ, USA, 2012. IEEE
  Press.

\bibitem{Gethers:ASE11}
{\sc Malcom Gethers, Huzefa Kagdi, Bogdan Dit, and Denys Poshyvanyk}.
\newblock An adaptive approach to impact analysis from change requests to
  source code.
\newblock In {\em Proceedings of the 2011 26th IEEE/ACM International
  Conference on Automated Software Engineering}, ASE '11, pages 540--543,
  Washington, DC, USA, 2011. IEEE Computer Society.

\bibitem{36Gu:ICSE10}
{\sc Zhongxian Gu, E.T. Barr, D.J. Hamilton, and Zhendong Su}.
\newblock Has the bug really been fixed?
\newblock In {\em Software Engineering, 2010 ACM/IEEE 32nd International
  Conference on}, volume~1, pages 55--64, May 2010.

\bibitem{19Guo:ICSE10}
{\sc Philip~J. Guo, Thomas Zimmermann, Nachiappan Nagappan, and Brendan
  Murphy}.
\newblock Characterizing and predicting which bugs get fixed: An empirical
  study of microsoft windows.
\newblock In {\em Proceedings of the 32Nd ACM/IEEE International Conference on
  Software Engineering - Volume 1}, ICSE '10, pages 495--504, New York, NY,
  USA, 2010. ACM.

\bibitem{Hossen:ICPC2014}
{\sc Md~Kamal Hossen, Huzefa Kagdi, and Denys Poshyvanyk}.
\newblock Amalgamating source code authors, maintainers, and change proneness
  to triage change requests.
\newblock In {\em Proceedings of the 22Nd International Conference on Program
  Comprehension}, ICPC 2014, pages 130--141, New York, NY, USA, 2014. ACM.

\bibitem{35Huo:ICSME14}
{\sc Da~Huo, Tao Ding, C.~McMillan, and M.~Gethers}.
\newblock An empirical study of the effects of expert knowledge on bug reports.
\newblock In {\em Software Maintenance and Evolution (ICSME), 2014 IEEE
  International Conference on}, pages 1--10, Sept 2014.

\bibitem{40Jeong:FSE2009}
{\sc Gaeul Jeong, Sunghun Kim, and Thomas Zimmermann}.
\newblock Improving bug triage with bug tossing graphs.
\newblock In {\em Proceedings of the the 7th Joint Meeting of the European
  Software Engineering Conference and the ACM SIGSOFT Symposium on The
  Foundations of Software Engineering}, ESEC/FSE '09, pages 111--120, New York,
  NY, USA, 2009. ACM.

\bibitem{18Jin:ICSE12}
{\sc Wei Jin and Alessandro Orso}.
\newblock Bugredux: Reproducing field failures for in-house debugging.
\newblock In {\em Proceedings of the 34th International Conference on Software
  Engineering}, ICSE '12, pages 474--484, Piscataway, NJ, USA, 2012. IEEE
  Press.

\bibitem{26Jin:ISSTA13}
{\sc Wei Jin and Alessandro Orso}.
\newblock F3: Fault localization for field failures.
\newblock In {\em Proceedings of the 2013 International Symposium on Software
  Testing and Analysis}, ISSTA 2013, pages 213--223, New York, NY, USA, 2013.
  ACM.

\bibitem{Kagdi:WCRE10}
{\sc H.~Kagdi, M.~Gethers, D.~Poshyvanyk, and M.L. Collard}.
\newblock Blending conceptual and evolutionary couplings to support change
  impact analysis in source code.
\newblock In {\em Reverse Engineering (WCRE), 2010 17th Working Conference on},
  pages 119--128, Oct 2010.

\bibitem{Kagdi:ICPC09}
{\sc H.~Kagdi and D.~Poshyvanyk}.
\newblock Who can help me with this change request?
\newblock In {\em Program Comprehension, 2009. ICPC '09. IEEE 17th
  International Conference on}, pages 273--277, May 2009.

\bibitem{Kagdi:ESE13}
{\sc Huzefa Kagdi, Malcom Gethers, and Denys Poshyvanyk}.
\newblock Integrating conceptual and logical couplings for change impact
  analysis in software.
\newblock {\em Empirical Software Engineering}, 18(5):933--969, 2013.

\bibitem{Huzefa:JSEP12}
{\sc Huzefa Kagdi, Malcom Gethers, Denys Poshyvanyk, and Maen Hammad}.
\newblock Assigning change requests to software developers.
\newblock {\em Journal of Software: Evolution and Process}, 24(1):3--33, 2012.

\bibitem{appendix}
{\sc Carlos Bernal~Cardenas Kevin~Moran, Mario Linares~Vasquez and Denys
  Poshyvanyk}.
\newblock Fusion online replication package
  \url{http://www.fusion-android.com}.

\bibitem{50Kifetew:ICST2014}
{\sc F.M. Kifetew, Wei Jin, R.~Tiella, A.~Orso, and P.~Tonella}.
\newblock Reproducing field failures for programs with complex grammar-based
  input.
\newblock In {\em Software Testing, Verification and Validation (ICST), 2014
  IEEE Seventh International Conference on}, pages 163--172, March 2014.

\bibitem{44Kim:TOSE2013}
{\sc Dongsun Kim, Yida Tao, Sunghun Kim, and A.~Zeller}.
\newblock Where should we fix this bug? a two-phase recommendation model.
\newblock {\em Software Engineering, IEEE Transactions on}, 39(11):1597--1610,
  Nov 2013.

\bibitem{46Kim:DSN2011}
{\sc Sunghun Kim, T.~Zimmermann, and N.~Nagappan}.
\newblock Crash graphs: An aggregated view of multiple crashes to improve crash
  triage.
\newblock In {\em Dependable Systems Networks (DSN), 2011 IEEE/IFIP 41st
  International Conference on}, pages 486--493, June 2011.

\bibitem{33Koru:IEEE2004}
{\sc A.~Gunes Koru and Jeff Tian}.
\newblock Defect handling in medium and large open source projects.
\newblock {\em IEEE Softw.}, 21(4):54--61, July 2004.

\bibitem{Linares-Vasquez:ICSM2012}
{\sc M.~Linares-Vasquez, K.~Hossen, Hoang Dang, H.~Kagdi, M.~Gethers, and
  D.~Poshyvanyk}.
\newblock Triaging incoming change requests: Bug or commit history, or code
  authorship?
\newblock In {\em Software Maintenance (ICSM), 2012 28th IEEE International
  Conference on}, pages 451--460, Sept 2012.

\bibitem{Linares:MSR15}
{\sc Mario Linares-V\'{a}squez, Martin White, Carlos Bernal-C\'{a}rdenas, Kevin
  Moran, and Denys Poshyvanyk}.
\newblock Mining android app usages for generating actionable gui-based
  execution scenarios.
\newblock In {\em 12th Working Conference on Mining Software Repositories
  (MSR'15)}, to appear, 2015.

\bibitem{Liu:ASE07}
{\sc Dapeng Liu, Andrian Marcus, Denys Poshyvanyk, and Vaclav Rajlich}.
\newblock Feature location via information retrieval based filtering of a
  single scenario execution trace.
\newblock In {\em Proceedings of the Twenty-second IEEE/ACM International
  Conference on Automated Software Engineering}, ASE '07, pages 234--243, New
  York, NY, USA, 2007. ACM.

\bibitem{Machiry:FSE2013}
{\sc Aravind Machiry, Rohan Tahiliani, and Mayur Naik}.
\newblock Dynodroid: An input generation system for android apps.
\newblock In {\em Proceedings of the 2013 9th Joint Meeting on Foundations of
  Software Engineering}, ESEC/FSE 2013, pages 224--234, New York, NY, USA,
  2013. ACM.

\bibitem{1Mani:FSE12}
{\sc Senthil Mani, Rose Catherine, Vibha~Singhal Sinha, and Avinava Dubey}.
\newblock Ausum: Approach for unsupervised bug report summarization.
\newblock In {\em Proceedings of the ACM SIGSOFT 20th International Symposium
  on the Foundations of Software Engineering}, FSE '12, pages 11:1--11:11, New
  York, NY, USA, 2012. ACM.

\bibitem{47Marsi:STVR2010}
{\sc Wes Masri}.
\newblock Fault localization based on information flow coverage.
\newblock {\em Software Testing, Verification and Reliability}, 20(2):121--147,
  2010.

\bibitem{Menzies:ICSM2008}
{\sc T.~Menzies and A.~Marcus}.
\newblock Automated severity assessment of software defect reports.
\newblock In {\em Software Maintenance, 2008. ICSM 2008. IEEE International
  Conference on}, pages 346--355, Sept 2008.

\bibitem{Moran:FSE15}
{\sc K.~Moran, M.~Linares-Vasquez, C.~Bernal-Cardenas, and D.~Poshyvanyk}.
\newblock Auto-completing bug reports for android applications.
\newblock In {\em Proceedings of 10th Joint Meeting of the European Software
  Engineering Conference and the 23rd ACM SIGSOFT Symposium on the Foundations
  of Software Engineering (ESEC/FSE?15)}, to appear, 2015.

\bibitem{Morville:04}
{\sc Peter Morville}.
\newblock User experience design.
  \url{http://semanticstudios.com/user_experience_design/}.

\bibitem{10Naguib:MSR13}
{\sc Hoda Naguib, Nitesh Narayan, Bernd Br\"{u}gge, and Dina Helal}.
\newblock Bug report assignee recommendation using activity profiles.
\newblock In {\em Proceedings of the 10th Working Conference on Mining Software
  Repositories}, MSR '13, pages 22--30, Piscataway, NJ, USA, 2013. IEEE Press.

\bibitem{14Nguyen:ASE12}
{\sc Anh~Tuan Nguyen, Tung~Thanh Nguyen, Tien~N. Nguyen, David Lo, and
  Chengnian Sun}.
\newblock Duplicate bug report detection with a combination of information
  retrieval and topic modeling.
\newblock In {\em Proceedings of the 27th IEEE/ACM International Conference on
  Automated Software Engineering}, ASE 2012, pages 70--79, New York, NY, USA,
  2012. ACM.

\bibitem{Nguyen:TSE14}
{\sc Bao Nguyen and Atif Memon}.
\newblock An observe-model-exercise* paradigm to test event-driven systems with
  undetermined input spaces.
\newblock {\em IEEE Transactions on Software Engineering}, 99(Preprints), 2014.

\bibitem{Nguyen:TOSE2014}
{\sc B.N. Nguyen and A.M. Memon}.
\newblock An observe-model-exercise; paradigm to test event-driven systems with
  undetermined input spaces.
\newblock {\em Software Engineering, IEEE Transactions on}, 40(3):216--234,
  March 2014.

\bibitem{48Podgurski:ICSE2003}
{\sc A.~Podgurski, D.~Leon, P.~Francis, W.~Masri, M.~Minch, Jiayang Sun, and
  Bin Wang}.
\newblock Automated support for classifying software failure reports.
\newblock In {\em Software Engineering, 2003. Proceedings. 25th International
  Conference on}, pages 465--475, May 2003.

\bibitem{Poshyvanyk:ICPC06}
{\sc D.~Poshyvanyk, Y.-G. Gueheneuc, A.~Marcus, G.~Antoniol, and V.~Rajlich}.
\newblock Combining probabilistic ranking and latent semantic indexing for
  feature identification.
\newblock In {\em Program Comprehension, 2006. ICPC 2006. 14th IEEE
  International Conference on}, pages 137--148, 2006.

\bibitem{Poshyvanyk:TSE07}
{\sc D.~Poshyvanyk, Y.-G. Gueheneuc, A.~Marcus, G.~Antoniol, and V.~Rajlich}.
\newblock Feature location using probabilistic ranking of methods based on
  execution scenarios and information retrieval.
\newblock {\em Software Engineering, IEEE Transactions on}, 33(6):420--432,
  June 2007.

\bibitem{Posyvanyk:ICPC07}
{\sc D.~Poshyvanyk and A.~Marcus}.
\newblock Combining formal concept analysis with information retrieval for
  concept location in source code.
\newblock In {\em Program Comprehension, 2007. ICPC '07. 15th IEEE
  International Conference on}, pages 37--48, June 2007.

\bibitem{Poshyvanyk:TSEM13}
{\sc Denys Poshyvanyk, Malcom Gethers, and Andrian Marcus}.
\newblock Concept location using formal concept analysis and information
  retrieval.
\newblock {\em ACM Trans. Softw. Eng. Methodol.}, 21(4):23:1--23:34, February
  2013.

\bibitem{Poshyvanyk:ESE09}
{\sc Denys Poshyvanyk, Andrian Marcus, Rudolf Ferenc, and Tibor Gyim\'{o}thy}.
\newblock Using information retrieval based coupling measures for impact
  analysis.
\newblock {\em Empirical Softw. Engg.}, 14(1):5--32, February 2009.

\bibitem{37Rahman:FSE2011}
{\sc Foyzur Rahman, Daryl Posnett, Abram Hindle, Earl Barr, and Premkumar
  Devanbu}.
\newblock Bugcache for inspections: Hit or miss?
\newblock In {\em Proceedings of the 19th ACM SIGSOFT Symposium and the 13th
  European Conference on Foundations of Software Engineering}, ESEC/FSE '11,
  pages 322--331, New York, NY, USA, 2011. ACM.

\bibitem{20Rastkar:ICSE10}
{\sc Sarah Rastkar, Gail~C. Murphy, and Gabriel Murray}.
\newblock Summarizing software artifacts: A case study of bug reports.
\newblock In {\em Proceedings of the 32Nd ACM/IEEE International Conference on
  Software Engineering - Volume 1}, ICSE '10, pages 505--514, New York, NY,
  USA, 2010. ACM.

\bibitem{Ravindranath:Mobisys2014}
{\sc Lenin Ravindranath, Suman Nath, Jitendra Padhye, and Hari Balakrishnan}.
\newblock Automatic and scalable fault detection for mobile applications.
\newblock In {\em Proceedings of the 12th Annual International Conference on
  Mobile Systems, Applications, and Services}, MobiSys '14, pages 190--203, New
  York, NY, USA, 2014. ACM.

\bibitem{Revelle:ICPC10}
{\sc M.~Revelle, B.~Dit, and D.~Poshyvanyk}.
\newblock Using data fusion and web mining to support feature location in
  software.
\newblock In {\em Program Comprehension (ICPC), 2010 IEEE 18th International
  Conference on}, pages 14--23, June 2010.

\bibitem{51Haihao:ICST2011}
{\sc Haihao Shen, Jianhong Fang, and Jianjun Zhao}.
\newblock Efindbugs: Effective error ranking for findbugs.
\newblock In {\em Software Testing, Verification and Validation (ICST), 2011
  IEEE Fourth International Conference on}, pages 299--308, March 2011.

\bibitem{7Shokripour:MSR13}
{\sc Ramin Shokripour, John Anvik, Zarinah~M. Kasirun, and Sima Zamani}.
\newblock Why so complicated? simple term filtering and weighting for
  location-based bug report assignment recommendation.
\newblock In {\em Proceedings of the 10th Working Conference on Mining Software
  Repositories}, MSR '13, pages 2--11, Piscataway, NJ, USA, 2013. IEEE Press.

\bibitem{22BugDigger}
{\sc BugsIO Solutions}.
\newblock Bugdigger.
\newblock http://bugdigger.com, December 2014.

\bibitem{Takala:ICST2011}
{\sc Tommi Takala, Mika Katara, and Julian Harty}.
\newblock Experiences of system-level model-based gui testing of an android
  application.
\newblock In {\em Proceedings of the 2011 Fourth IEEE International Conference
  on Software Testing, Verification and Validation}, ICST '11, pages 377--386,
  Washington, DC, USA, 2011. IEEE Computer Society.

\bibitem{25Tassey:NIST}
{\sc G.~Tassey}.
\newblock The economic impacts of inadequate infrastructure for software
  testing.
\newblock Technical report, National Institute of Standards and Technology,
  2002.

\bibitem{39Vidacs:CSMR14}
{\sc L.~Vidacs, A.~Beszedes, D.~Tengeri, I.~Siket, and T.~Gyimothy}.
\newblock Test suite reduction for fault detection and localization: A combined
  approach.
\newblock In {\em Software Maintenance, Reengineering and Reverse Engineering
  (CSMR-WCRE), 2014 Software Evolution Week - IEEE Conference on}, pages
  204--213, Feb 2014.

\bibitem{13Wang:ICPC14}
{\sc Shaowei Wang and David Lo}.
\newblock Version history, similar report, and structure: Putting them together
  for improved bug localization.
\newblock In {\em Proceedings of the 22Nd International Conference on Program
  Comprehension}, ICPC 2014, pages 53--63, New York, NY, USA, 2014. ACM.

\bibitem{17Wang:ICSE08}
{\sc Xiaoyin Wang, Lu~Zhang, Tao Xie, John Anvik, and Jiasu Sun}.
\newblock An approach to detecting duplicate bug reports using natural language
  and execution information.
\newblock In {\em Proceedings of the 30th International Conference on Software
  Engineering}, ICSE '08, pages 461--470, New York, NY, USA, 2008. ACM.

\bibitem{53Weiss:MSR2007}
{\sc Cathrin Weiss, Rahul Premraj, Thomas Zimmermann, and Andreas Zeller}.
\newblock How long will it take to fix this bug?
\newblock In {\em Proceedings of the Fourth International Workshop on Mining
  Software Repositories}, MSR '07, pages 1--, Washington, DC, USA, 2007. IEEE
  Computer Society.

\bibitem{45Park:AAAI2011}
{\sc Jin woo Park, Mu-Woong Lee, Jinhan Kim, Seung won Hwang, and Sunghun Kim}.
\newblock Costriage: A cost-aware triage algorithm for bug reporting systems,
  2011.

\bibitem{42Wu:ISSTA2014}
{\sc Rongxin Wu, Hongyu Zhang, Shing-Chi Cheung, and Sunghun Kim}.
\newblock Crashlocator: Locating crashing faults based on crash stacks.
\newblock In {\em Proceedings of the 2014 International Symposium on Software
  Testing and Analysis}, ISSTA 2014, pages 204--214, New York, NY, USA, 2014.
  ACM.

\bibitem{21Zhou:CIKM12}
{\sc Jian Zhou and Hongyu Zhang}.
\newblock Learning to rank duplicate bug reports.
\newblock In {\em Proceedings of the 21st ACM International Conference on
  Information and Knowledge Management}, CIKM '12, pages 852--861, New York,
  NY, USA, 2012. ACM.

\bibitem{8Zhou:ICSE12}
{\sc Jian Zhou, Hongyu Zhang, and David Lo}.
\newblock Where should the bugs be fixed? - more accurate information
  retrieval-based bug localization based on bug reports.
\newblock In {\em Proceedings of the 34th International Conference on Software
  Engineering}, ICSE '12, pages 14--24, Piscataway, NJ, USA, 2012. IEEE Press.

\end{thebibliography}
\end{document}